\DocumentMetadata{}
\documentclass[nonacm,sigplan]{acmart}

\usepackage[]{hyperref}
\usepackage{multirow}
\usepackage{listings}
\usepackage{subcaption}
\usepackage{tabularx}
\usepackage{adjustbox}
\usepackage{threeparttable}
\usepackage{pifont}
\usepackage{makecell}
\usepackage{longtable}
\usepackage{colortbl}
\usepackage{rotating}

\lstdefinestyle{mystyle}{
    language=C++,
    backgroundcolor=\color{white},
    basicstyle=\ttfamily\small,
    keywordstyle=\color{blue}\bfseries,
    commentstyle=\color{gray}\itshape,
    stringstyle=\color{purple},
    numberstyle=\tiny\color{gray},
    numbers=left,
    numbersep=10pt,
    tabsize=2,
    breaklines=true,
    breakatwhitespace=false,
    showstringspaces=false,
    columns=fullflexible,
    keepspaces=true,
    frame=single,
    framerule=0.5pt,
    rulecolor=\color{black},
    captionpos=b
}
\newcommand{\cmark}{\ding{51}}
\newcommand{\xmark}{\ding{55}}
\definecolor{lightred}{rgb}{1,0.8,0.8}
\definecolor{lightgreen}{rgb}{0.8,1,0.8}

\begin{document}

\title{Voyager: An End-to-End Framework for Design-Space Exploration and Generation of DNN Accelerators}

\author{Kartik Prabhu}
    \affiliation{%
      \institution{Stanford University}
      \city{Stanford}
      \country{USA}}
    \email{kprabhu7@stanford.edu}

\author{Jeffrey Yu}
    \affiliation{%
      \institution{Stanford University}
      \city{Stanford}
      \country{USA}}
    \email{jeffreyy@stanford.edu}

\author{Xinyuan Allen Pan}
    \affiliation{%
      \institution{Stanford University}
      \city{Stanford}
      \country{USA}}
    \email{allpan@stanford.edu}

\author{Zhouhua Xie}
    \affiliation{%
      \institution{Stanford University}
      \city{Stanford}
      \country{USA}}
    \email{xzh015@stanford.edu}

\author{Abigail Aleshire}
    \affiliation{%
      \institution{Stanford University}
      \city{Stanford}
      \country{USA}}
    \email{abbya@stanford.edu}

\author{Zihan Chen}
    \affiliation{%
      \institution{Stanford University}
      \city{Stanford}
      \country{USA}}
    \email{zihan410@stanford.edu}

\author{Ammar Ali Ratnani}
    \affiliation{%
      \institution{Stanford University}
      \city{Stanford}
      \country{USA}}
    \email{aratnani@stanford.edu}

\author{Priyanka Raina}
    \affiliation{%
      \institution{Stanford University}
      \city{Stanford}
      \country{USA}}
    \email{praina@stanford.edu}

\begin{abstract}

While deep neural networks (DNNs) have achieved state-of-the-art performance in fields from computer vision to natural language processing, efficiently running these computationally demanding models requires specialized hardware accelerators. However, designing these accelerators is a time-consuming, labor-intensive process that does not scale well across multiple design points. While prior efforts have sought to automate DNN accelerator generation, they typically offer limited parameterization, cannot produce high-performance, tapeout-ready designs, provide limited support for multiple datatypes and quantization schemes, and lack an integrated, end-to-end software compiler.

This work proposes Voyager, a high-level synthesis (HLS)-based framework for rapid design space exploration and generation of DNN accelerators. Voyager overcomes the limitations of prior work by offering extensive configurability across technology nodes, clock frequencies, and scales, with customizable parameters such as number of processing elements, on-chip buffer sizes, and external memory bandwidth. Voyager supports a much wider variety of datatypes and quantization schemes versus prior work, including both built-in arbitrary-length floating-point, posit and integer formats, as well as user-defined custom formats with both per-tensor scaling and microscaling quantization. Voyager's PyTorch-based compiler efficiently maps neural networks end-to-end on the generated hardware, with support for quantization, operation fusion, and tiling.

We evaluate Voyager on state-of-the-art vision and language models. Voyager enables fast design-space exploration with full-dataset accuracy evaluation for different datatypes and quantization schemes. Generated designs achieve a high utilization across models and scales, up to 99.8\%, and outperform prior generators with up to 61\% lower latency and 56\% lower area. Compared to hand-optimized accelerators, Voyager achieves comparable performance, while offering much greater automation in design and workload mapping.

\end{abstract}

\maketitle % should come after the abstract
\pagestyle{plain} % should come right after \maketitle

\section{Introduction}
Deep neural networks (DNNs) are highly successful across many domains, but require substantial compute and memory resources. While many DNN accelerators have been proposed to address these challenges, designing efficient accelerators is an engineering-intensive process. As neural network models evolve rapidly, developing custom accelerators for each new type of model is increasingly impractical.

To address this, several DNN accelerator generators~\cite{magnet,gemmini-dac,nvdla} have been proposed. While these frameworks help automate hardware generation, they often fall short in four key areas: (1) lack of support for comprehensive design space exploration, (2) inability to produce tapeout-ready systems, (3) lack of support for quantization schemes, such as micro-scaling, and (4) lack of a robust compiler to map high-level DNN models to the generated accelerators.

\begin{figure}
    \centering
    \includegraphics[width=1\linewidth]{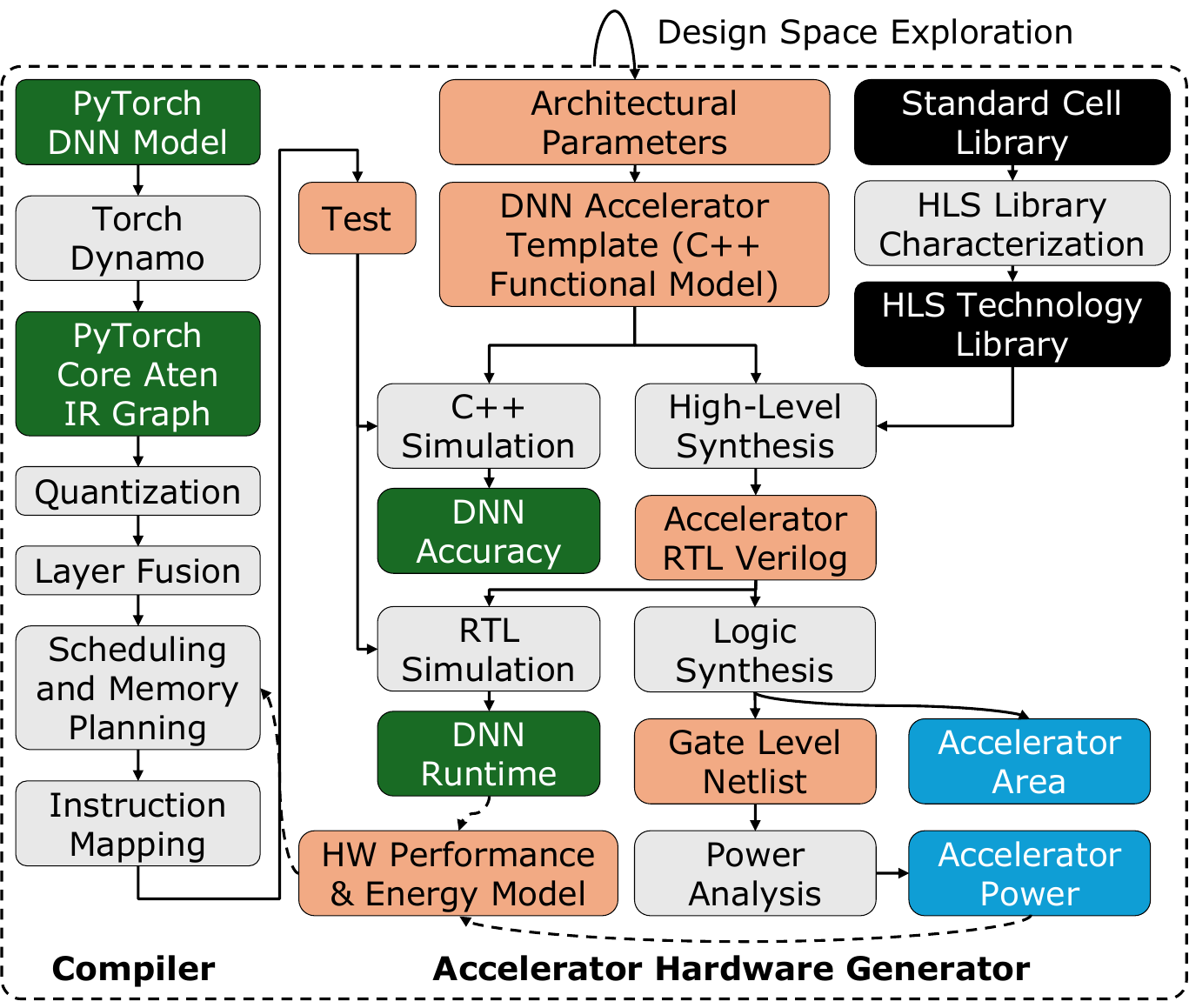}
    \caption{Voyager is an end-to-end framework for generating and evaluating neural network accelerators. It uses a high-level synthesis flow built around a templated accelerator design. Given architectural parameters, Voyager generates RTL implementations targeting a specified clock frequency and technology node. The C++ model enables fast evaluation of quantized DNN accuracy, while RTL simulation and synthesis provide runtime, energy, and area estimates.}
    \label{fig:overview}
\end{figure}

In this paper, we present Voyager, an end-to-end high-level synthesis (HLS)-based framework for design space exploration and generation of neural network accelerators. As shown in \autoref{fig:overview}, given architectural parameters such as data types, scaling factor type and granularity, compute parallelization, and buffer sizes, our framework produces performant, tapeout-ready register-transfer level (RTL) accelerator implementations. It also includes a co-designed compiler that efficiently maps PyTorch models to the generated hardware.

This paper makes the following contributions:
\begin{itemize}
    \item We present an end-to-end high-level synthesis (HLS)-based framework for design-space exploration and generation of neural network accelerators. Our framework generates accelerators that are optimized for a given set of machine learning models, technology node, and performance constraints. The design-space exploration tool supports accelerators with different numbers of processing elements, on-chip buffer sizes, and off-chip memory bandwidth, and generates designs with a highly-configurable dataflow, with programmable loop tiling, ordering, unrolling and fusion.
    \item Our generator supports a wider range of datatypes and quantization schemes compared to prior work. In addition to floating-point, posit, and integer datatypes with coarse-grained scaling factors supported by prior work, we additionally can generate accelerators with mixed-precision, multi-precision and custom user-defined types. Furthermore, we support any of these types with microscaling. Using our system, we can evaluate the costs and benefits of microscaling compared to coarser-grained scaling techniques. As opposed to prior work on microscaling that only evaluates accuracy using fake-quantization, our system leverages HLS to both generate hardware and provide fast quantized inference accuracy results that are bit-accurate with hardware. Particularly, our system accurately models the datatypes used for accumulation and non-linearities, unlike fake-quantization.
    \item Finally, we present a compiler that maps PyTorch models to any instance of the generated accelerator. Our compiler supports quantization, hardware-aware operation fusion, and we integrate with our extended version of Interstellar \cite{interstellar} to find the optimal scheduling (loop tiling, ordering, unrolling and fusion) of a neural network on the accelerator.
\end{itemize}

We evaluate Voyager by generating accelerators that efficiently execute a wide variety of models used for image classification, language understanding and generation. Our results show that microscaled integers achieve inference accuracy that is within 1\% of 32-bit floating point, but with up to 23\% lower area than low-precision floating point formats. Voyager generated designs achieve up to 56\% lower area and 61\% lower runtime compared to prior generators, like Gemmini and NVDLA at iso-hardware configuration. Compared to hand-optimized accelerators, we achieve up to 48\% better performance, while providing greater design automation.

\section{Related Work}

\begin{table*}
\centering

\begin{tabular}{|l|c|c|c|c|c|c|c|}
\hline
\multirow{1}{*}{\textbf{Feature}} & \textbf{Tandem} \cite{tandem_processor}& \textbf{Builder} \cite{dnnbuilder} & \textbf{MAERI} \cite{maeri} & \textbf{NVDLA} \cite{nvdla} & \textbf{MAGNet} \cite{magnet}& \textbf{Gemmini} \cite{gemmini-dac} & \textbf{Voyager} \\
\hline
Datatypes & Unspecified & Unspecified & Unspecified & Int/Float & Int & Int/Float & \shortstack{Int/Float/\\Posit/Custom} \\
\hline
Multi-Precision & \cellcolor{lightred}\xmark & \cellcolor{lightred}\xmark & \cellcolor{lightred}\xmark & \cellcolor{lightred}\xmark & \cellcolor{lightred}\xmark & \cellcolor{lightred}\xmark & \cellcolor{lightgreen}\cmark \\
\hline
Microscaling & \cellcolor{lightred}\xmark & \cellcolor{lightred}\xmark & \cellcolor{lightred}\xmark & \cellcolor{lightred}\xmark & \cellcolor{lightred}\xmark & \cellcolor{lightred}\xmark & \cellcolor{lightgreen}\cmark \\
\hline
Abstraction& \makecell{RTL Verilog} & \makecell{RTL } & \makecell{RTL Verilog} & \makecell{RTL Verilog} & \makecell{HLS C++} & \makecell{RTL Chisel} & \makecell{HLS C++} \\
\hline
SW Support & ONNX & Custom & Custom & Caffe & Custom & ONNX/C & PyTorch \\
\hline
SoC Integration & \cellcolor{lightred}\xmark & \cellcolor{lightred}\xmark & \cellcolor{lightred}\xmark & \cellcolor{lightred}\xmark & \cellcolor{lightred}\xmark & \cellcolor{lightgreen}\cmark & \cellcolor{lightgreen}\cmark \\
\hline
\end{tabular}
    \caption{Comparison with prior DNN accelerator generators.}
    \label{tab:related-work}
\end{table*}

Exploring the design space of DNN accelerators has been a focus of several prior works, which rely on architectural modeling. Tools such as Interstellar~\cite{interstellar}, Timeloop~\cite{timeloop}, MAESTRO~\cite{maestro}, and ZigZag~\cite{zigzag} calculate workload costs by generating mappings and evaluating them using an analytical cost model. Interstellar~\cite{interstellar} estimates energy with a coarse-grained model of memory access and compute energy, using heuristics to prune the search space. Timeloop~\cite{timeloop} adopts a similar approach, but supports exhaustive or random-sampling search. MAESTRO~\cite{maestro} focuses on scheduling search for a fixed dataflow. ZigZag~\cite{zigzag} extends these  works by offering greater flexibility in the architectural template and expanding the space of valid mappings. While useful for design-space exploration, these tools do not generate hardware.
 
Beyond modeling, several works generate synthesizable hardware for DNN accelerators, including Tandem Processor~\cite{tandem_processor}, DNNBuilder~\cite{dnnbuilder}, MAERI~\cite{maeri}, NVDLA~\cite{nvdla}, MAGNet~\cite{magnet}, and Gemmini~\cite{gemmini-dac}. These works differ in architecture, configurability, and software support.  

Tandem Processor~\cite{tandem_processor} targets non-GEMM operations and enables end-to-end execution of DNN models when combined with a GEMM unit. It uses software-managed scratchpads and a custom ISA to avoid the overheads of address calculation and loop control as in conventional processors. A custom compiler lowers ONNX graphs to the ISA. 

DNNBuilder~\cite{dnnbuilder} generates designs for FPGAs with a pipelined architecture, mapping each pipeline stage to a neural network layer. The design is fixed to a single model, unlike other generators, including Voyager, that produce more general-purpose accelerators capable of supporting multiple DNN models.

MAERI~\cite{maeri} builds accelerators with a flexible interconnection fabric. The architecture consists of MAC units connected via a tree-based interconnection that supports a wide range of dataflows. MAERI can generate designs of various scales, with configurable numbers of MAC units and SRAM sizes. However, it lacks a strong software stack.

NVDLA~\cite{nvdla} supports a wide range of DNN operations via specialized blocks for convolution, activation functions, pooling, normalization, and reshaping. It offers configurability in datatypes, MAC array sizes, memory interface widths, and buffer sizes. Each specialized block can be independently sized and selectively included or excluded. Its software stack compiles Caffe models to hardware instructions.

MAGNet~\cite{magnet} uses an HLS template-based approach, with an array of PEs connected to a global buffer. Each PE integrates vector MAC units, local buffers, and post-processing units for ReLU, pooling, scaling, and quantization. It is configurable along several dimensions, including the number of PEs, buffer sizes, number of vector lanes, vector widths, and integer precision. However, MAGNet lacks a software stack and requires manual mapping of neural network layers.

Gemmini~\cite{gemmini-dac} is a Chisel-based generator of designs with PEs, scratchpad and accumulator memories, and additional units for pooling and activation functions. Gemmini can generate both systolic arrays and vector engines, with configurability over array sizes, buffer sizes, and the inclusion of optional blocks such as im2col, pooling, or transpose units. Gemmini also supports both arbitrary precision integer and floating-point data types. Its software stack adopts the ONNX format as the model specification. 

In contrast to these works, Voyager generates accelerators that can accelerate neural networks end-to-end with far greater flexibility. Voyager supports a range of emerging datatypes, as well as advanced quantization strategies such as microscaling, which are increasingly critical for sub-8-bit precision. Prior generators also lack a robust software stack, requiring manual effort in converting models into custom formats for mapping them to hardware. In contrast, Voyager's PyTorch-based compiler enables automatic quantization, fusion, and scheduling, enabling efficient execution on generated accelerators with minimal manual effort.

 \section{Voyager's Accelerator Template}
\begin{figure*}[t]
\centering
\includegraphics[width=1\linewidth]{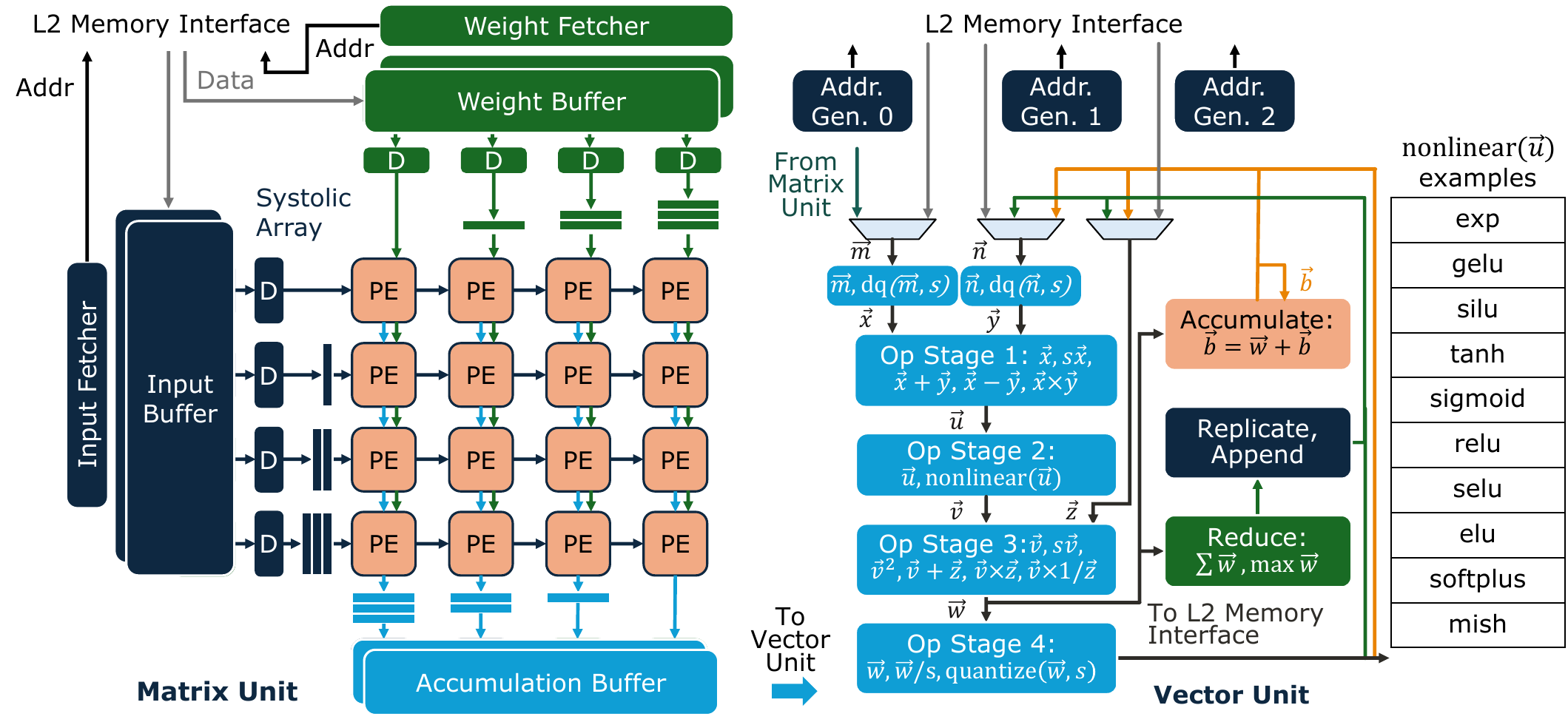}
\caption{Voyager's accelerator template has a matrix and a vector unit. Matrix unit has input and weight fetchers and buffers, an accumulation buffer, and a systolic array. Vector unit is a multi-stage pipeline for elementwise and reduction operations.}
\label{fig:accelerator}
\end{figure*}

Voyager produces hardware instances from a flexible accelerator template, illustrated in \autoref{fig:accelerator}. The template has a matrix unit for convolution and general matrix multiplication (GEMM), and a programmable vector unit for non-GEMM operations. Both units have custom instruction sets.

\subsection{Matrix Unit}
The matrix unit consists of a weight-stationary systolic array, where each processing element (PE) performs a multiply-accumulate (MAC) operation. Input channels are unrolled vertically, and output channels are unrolled horizontally. Inputs activations flow from left to right, while partial sums propagate from top to bottom. Weights are preloaded into PEs and remain stationary throughout a tile's computation.
 
To maintain high utilization and avoid stalls from weight loading, each PE has triple-buffered registers interconnected by a column-wise shift-register chain. While one tile is being processed, weights for the next tile can be shifted in concurrently, enabling tile-level pipelining. Partial sums exiting the array are optionally combined with bias terms and written into a dedicated accumulation buffer.

The matrix unit supports runtime-configurable dataflow through programmable address generators that fetch tiles of inputs and weights from main memory and store them in independent double buffers, overlapping memory accesses with computation for higher throughput. These address generators have fine-grained control of loop ordering and bounds at both the L1 and L2 memory levels. Additionally, the input and weight controllers support in-place transpose and permute operations via register buffers, enabling fusion of transpose with GEMM operations.

\subsection{Vector Unit}

The \(N\)-wide multi-stage vector unit executes non-linear activations (e.g., ReLU, GeLU), element-wise operations (e.g., residual additional), complex functions such as softmax and layer normalization, and quantization/dequantization. It includes a multi-stage operation pipeline, a reduction unit for summation and max reduction, and an accumulator. Additionally, the unit includes three independent address generators that can access main memory. 

To minimize memory accesses, outputs from the matrix unit feed directly into the vector unit, enabling fusion of matrix and vector operations. However, this tight coupling can stall the matrix unit if the vector unit is bottlenecked by memory access and exerts backpressure. To mitigate this, Voyager can generate designs with a double-buffered accumulation buffer: matrix unit outputs are instead stored in the buffer, and the vector unit independently reads from it while the matrix unit processes the next tile.

Within the operation pipeline, shown in \autoref{fig:accelerator} on the right, each stage performs an element-wise operation or forwards data unmodified. To select operations for each stage, we first construct a dataflow graph of core operations commonly used in DNNs, such as layer normalization and softmax. By merging several dataflow graphs, we create a complete pipeline that maximizes the reuse of arithmetic operators without impacting latency. Outputs from the vector pipeline can be routed to memory, the accumulator, or the reduction unit. The accumulator and reducer outputs can be fed back into the pipeline as operands for subsequent operations, or written out to memory.

Nonlinear functions are supported through a configurable polynomial approximation block. Rather than using lookup tables or fixed polynomial circuits for each function, we create a single programmable unit. At compile time, each nonlinear function in the model is approximated with a free-knot spline of seven quadratic segments. Segment boundaries and coefficients are encoded into the instruction. At runtime, the unit selects the segment based on the input, and evaluates the polynomial, providing accurate, hardware-efficient support for nonlinearities.

\begin{figure}[t!]
    \centering
    \includegraphics[width=0.9\linewidth]{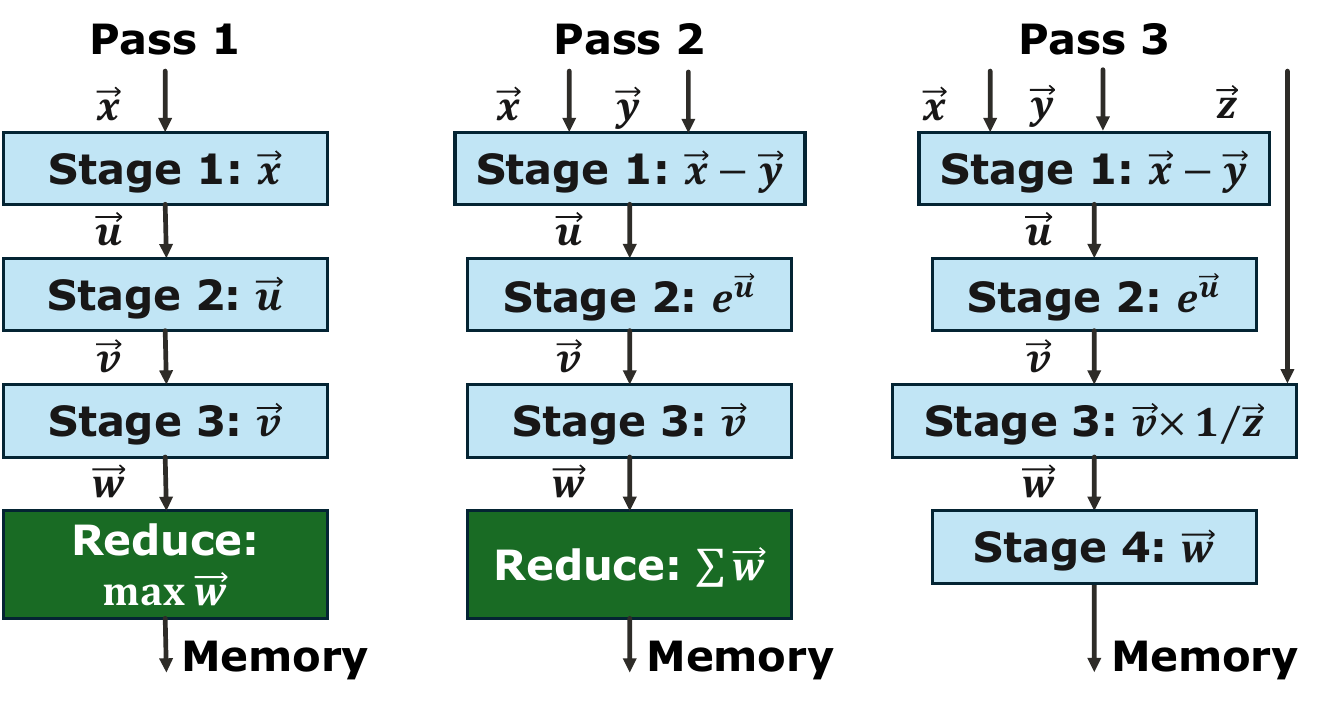}
    \caption{Three-pass mapping of softmax on vector unit.}
    \label{fig:softmax_mapping}
\end{figure}

To illustrate the vector unit's operation, \autoref{fig:softmax_mapping} shows how softmax is mapped in three passes. First, the tensor is streamed from memory to the reduction unit to compute the maximum value across the reduction dimension, and then written back to memory. Next, both the original tensor and the maximum are read from memory. Each element is shifted by subtracting the maximum, exponentiated, and reduced to compute the normalization denominator, which is written back to memory. Finally, the original tensor and max and sum are read from memory, and the vector pipeline performs the max subtraction, exponentiation, and division by the sum to produce the final probabilities, which are written to memory.

\subsection{User-Defined Extensions}

\begin{figure}[b!]
    \centering
    \includegraphics[width=\linewidth]{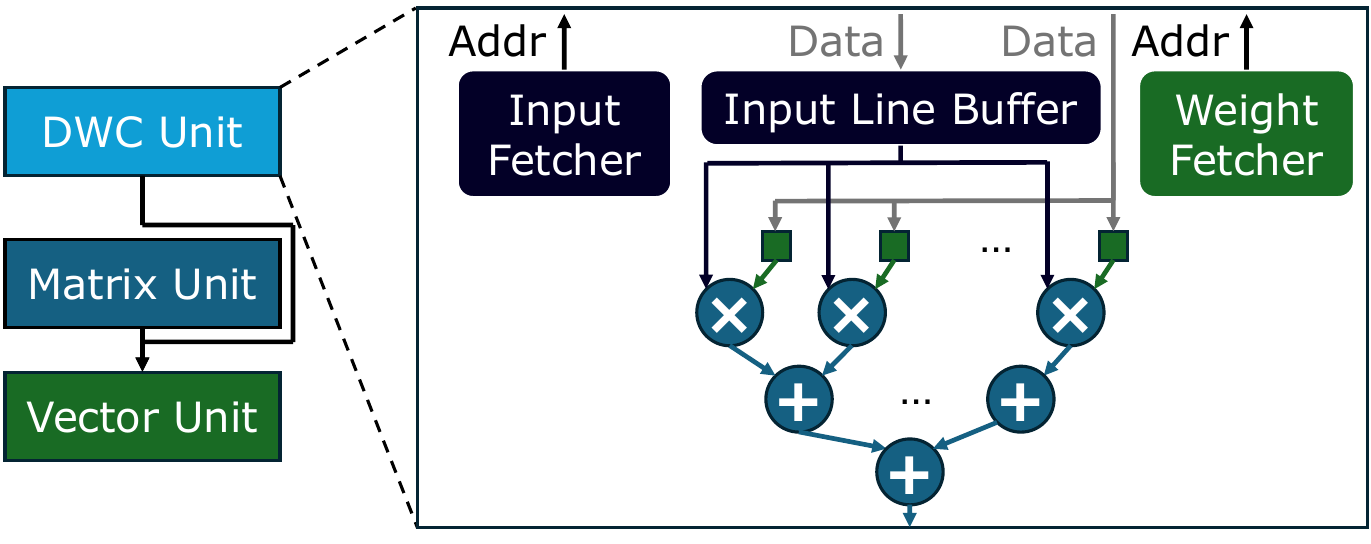}
    \caption{Example of a user-defined extension: a depthwise convolution unit with configurable MAC trees. Outputs flow directly into the vector unit for further processing.}
    \label{fig:dwc_unit}
\end{figure}

Voyager supports user-defined compute blocks to specialize the accelerator for operations not efficiently handled by the matrix or vector units. For example, depthwise convolutions (DWC) exhibit limited data reuse compared to standard convolutions, making systolic array-based acceleration less effective. To address this, a user can integrate a dedicated DWC unit, as shown in \autoref{fig:dwc_unit}. The block consists of input and weight fetchers, input line buffers, and a configurable number of MAC trees, each designed to reduce a $3\times3$ filter. For larger filter sizes, the compiler decomposes them into multiple $3\times3$ filters through slicing and padding, with the final accumulation performed in the vector unit. This keeps the hardware optimized for $3\times3$ filters, which dominate most models, while relying on the compiler to support other sizes. 

By supporting such extensions, Voyager integrates specialized blocks seamlessly with its baseline architecture, broadening the range of workloads it can efficiently accelerate.

\subsection{SoC Integration}

\begin{figure}
    \centering
    \includegraphics[width=\linewidth]{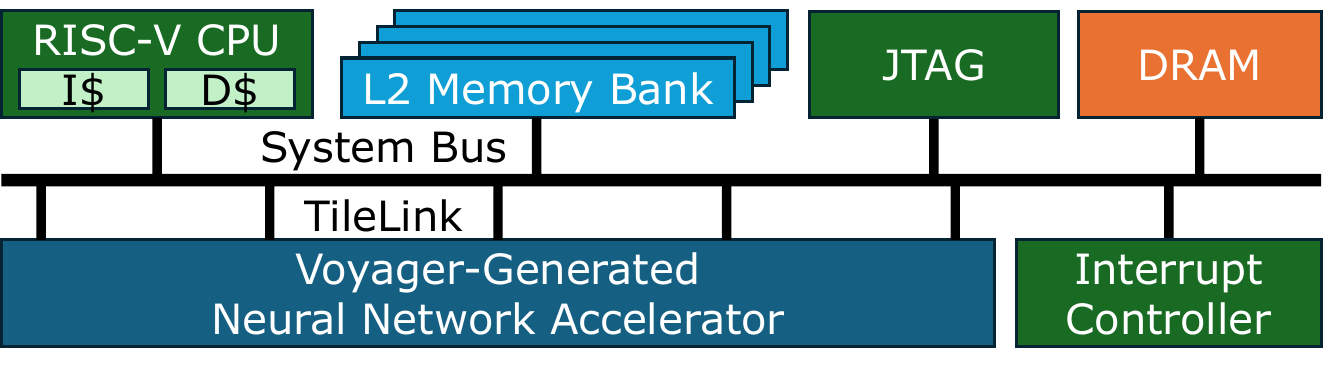}
    \caption{Voyager-generated accelerators can be integrated into an SoC using the Chipyard framework.}
    \label{fig:soc-integration}
\end{figure}

Voyager-generated accelerators can be integrated into system-on-chip (SoC) platforms using the Chipyard~\cite{chipyard} framework, as shown in \autoref{fig:soc-integration}. Chipyard provides a complete infrastructure for building RISC-V SoCs, such as processor cores, TileLink-based interconnect generators, memory subsystems, and peripheral devices.

We leverage Chipyard's support for custom accelerators by integrating Voyager-generated designs as memory-mapped I/O peripherals. The host CPU communicates with the accelerator through memory-mapped control registers, which are used to configure operations on the matrix and vector units. For memory accesses, the accelerator uses  dedicated lightweight TileLink ports to connect to the system bus, enabling seamless access to on-chip L2 memory and off-chip DRAM.

This integration demonstrates that Voyager-generated designs are not only effective as standalone accelerators, but also compatible with state-of-the-art SoC development flows, supporting both standalone and full-system evaluation.

\section{Hardware Generator Parameterization}
\begin{figure}[t]
\centering
\includegraphics[width=\linewidth]{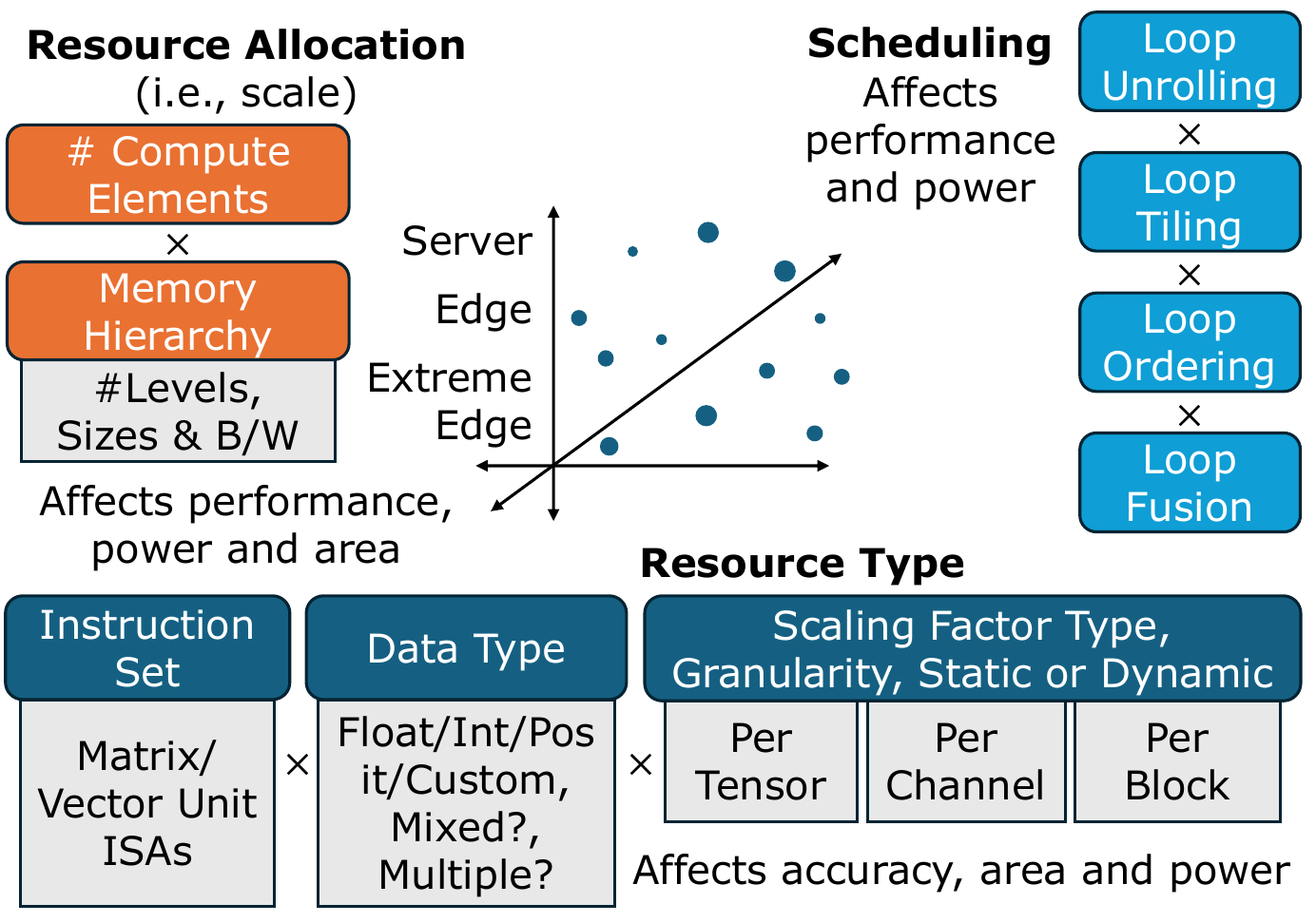}
\caption{DNN accelerator design space.}
\label{fig:dse}
\end{figure}

The Voyager generator exposes a large design space to the users, enabling them to generate designs that vary across three axes (\autoref{fig:dse}): resource allocation, resource type, and scheduling. Each axis captures a different aspect of the hardware (and software) design space and exposes tunable parameters that influence performance, power, area, and accuracy. To support this high degree of flexibility without sacrificing hardware efficiency, Voyager makes use of C++ metaprogramming features. Since HLS tools operate exclusively at compile-time, Voyager only utilizes the C++ features that enable compile-time specialization, particularly templates. 

\subsection{Resource Allocation: Compute/Memory Scaling}
Voyager can generate designs of different sizes by configuring the number of compute elements and memory resources. For example, the dimensions of the systolic array are configurable to match the target workloads' compute requirements, while the input, weight, and accumulation buffer sizes can be adjusted independently. Memory bandwidth to the main memory is also configurable. These parameters directly affect performance, area, and power and are chosen based on constraints such as latency targets or energy budgets. Larger systolic arrays reduce latency but incur higher area and power costs. Larger on-chip buffers enable greater data reuse and reduce main memory accesses, but increase buffer access energy and area \cite{interstellar}. These tradeoffs play a key role in design-space exploration and must be carefully balanced to meet system-level targets and constraints.

\subsection{Resource Type: Datatypes and Quantization}
Voyager can generate accelerators that support a variety of datatypes and quantization schemes, enabling users to generate accelerator variants tailored for specific classes of workloads. For instance, users can generate a standard BFloat16 accelerator, an INT8 accelerator with per-channel scaling, or an FP6 accelerator with microscaling. 

\paragraph{Datatype Parameterization} Voyager accomplishes hardware datatype parameterization through C++ templates. Templates in C++ enable generic programming, allowing developers to define functions and classes parameterized on type. During compilation, the compiler performs type substitution and generates specialized versions of these generic functions and classes with concrete types.

As shown in \autoref{lst:pe_generic}, the \texttt{ProcessingElement} module is the compute element of the systolic array, and performs a MAC operation on three operands. This module is templated on three datatypes: \texttt{InputType} for inputs, \texttt{WeightType} for weights, and \texttt{PsumType} for partial sums. The \texttt{SystolicArray} module then instantiates an array of \texttt{ProcessingElement}s, with concrete types for inputs, weights, and partial sums.

\begin{lstlisting}[language=C++, frame=none, caption={Generic processing element module.}, label={lst:pe_generic},float]
template<typename InputType, typename WeightType,
         typename PsumType>
SC_MODULE(ProcessingElement) {
  ...
  void compute(InputType input, WeightType weight, PsumType& psum){
    psum = input.mac(weight, psum);
  }};
SC_MODULE(SystolicArray){
  ...
  ProcessingElement<int8, int8, int32> pe[16][16];
};
\end{lstlisting}

Voyager provides three datatype families by default: integer, floating-point, and posit \cite{posit}, each implementing arithmetic and logical function according to its own semantics. These classes are also templated on configuration parameters, such as bit width, exponent size, or signedness (\autoref{lst:pe_variants}). Each type implements a \texttt{mac} function, enabling seamless type substitution during compilation with C++ templates.

\begin{lstlisting}[language=C++, frame=none, caption={Datatype class implementations.}, label={lst:pe_variants},float]
template<int Width, bool Signed>
class Int {
  ac_int<Width, Signed> value;
  Int mac(Int weight, Int psum) {
    return value * weight.value + psum.value; 
  }
};
template<int MantissaBits, int ExponentBits>
class FP {
  ac_std_float<MantissaBits, ExponentBits> value;
  FP mac(FP weight, FP psum) {
    return value.fma(weight, psum);
  }
};
template<int Width, int EsBits>
class Posit {
  ac_int<Width, false> bits;
  Posit mac(Posit weight, Posit psum) {
    ...
  }
};
\end{lstlisting}

In addition to the built-in types, Voyager supports extensibility through user-defined datatypes. To add a new type, users simply define a class that implements the expected interface, including the \texttt{mac} function and other arithmetic and logical operations. This mechanism is flexible enough to support configurations such as vectorized PEs (\autoref{lst:vector_type}) and unconventional formats like a codebook-based NormalFloat4 \cite{nf4} (\autoref{lst:nf4}).

\begin{lstlisting}[language=C++, frame=none, caption={Voyager supports vectorized PEs.}, label={lst:vector_type},float]
template<typename T, int Width>
class Vector {
  T data[Width];
  Vector mac(Vector weight, Vector psum) {
    for(int i = 0; i < Width; i++) { 
      data[i].mac(weight[i], psum[i]);
    }}};
\end{lstlisting}

\begin{lstlisting}[language=C++, frame=none, caption={Voyager supports novel datatypes.}, label={lst:nf4},float]
class NormalFloat4 {
  ac_int<4, false> bits;
  static const FP codebook[16];
  NormalFloat4 mac(NormalFloat4 weight, NormalFloat4 psum) {
    for(int i = 0; i < Width; i++) { 
      codebook[bits].mac(codebook[weight.bits], codebook[psum.bits]);
    }}};
\end{lstlisting}

Voyager also supports multi-precision designs, where the hardware can operate on different datatypes that are selectable at run-time. For example, some layers in a model are more sensitive to quantization errors and may require higher precision, while others can tolerate lower precision to improve efficiency. 

To enable this flexibility, Voyager uses variadic C++ templates and partial template specialization to generate hardware modules that can seamlessly support multiple datatypes within the same design. For example, the \texttt{InputController} fetches tiles of inputs from L2 memory. In order to support fetching multiple datatypes, this module is instantiated with tuples of supported types, as shown in \autoref{lst:mixed-precision}. This approach generates datatype-specific fetch logic that can be selectively used at run-time. The compiler encodes the precision required for each layer into the instruction, enabling the hardware to flexibly support multiple precisions.

\begin{lstlisting}[language=C++, frame=none, caption={The InputController template supports multiple datatypes through variadic templates using tuples.}, label={lst:mixed-precision},float]
template <typename InputTypes>
struct InputController;

template <typename... InputTypes>
struct InputController<std::tuple<InputTypes...>>{...};

// InputController instantiated with a single type
InputController<std::tuple<FP8>> inputController;
// or multiple types
InputController<std::tuple<FP16,FP8,FP4>> inputController;
\end{lstlisting}

\paragraph{Quantization Scheme Parameterization}
To support a wide range of accuracy-efficiency tradeoffs, Voyager can generate accelerators with various quantization schemes, including per-tensor and per-channel quantization. To support these schemes, Voyager-generated accelerators consist of a quantization stage at the output of the vector unit as well as dequantization hardware within the vector unit's fetcher, allowing for fusion of quantization and dequantization with the previous operation.

\begin{figure}
    \centering
    \includegraphics[width=\linewidth]{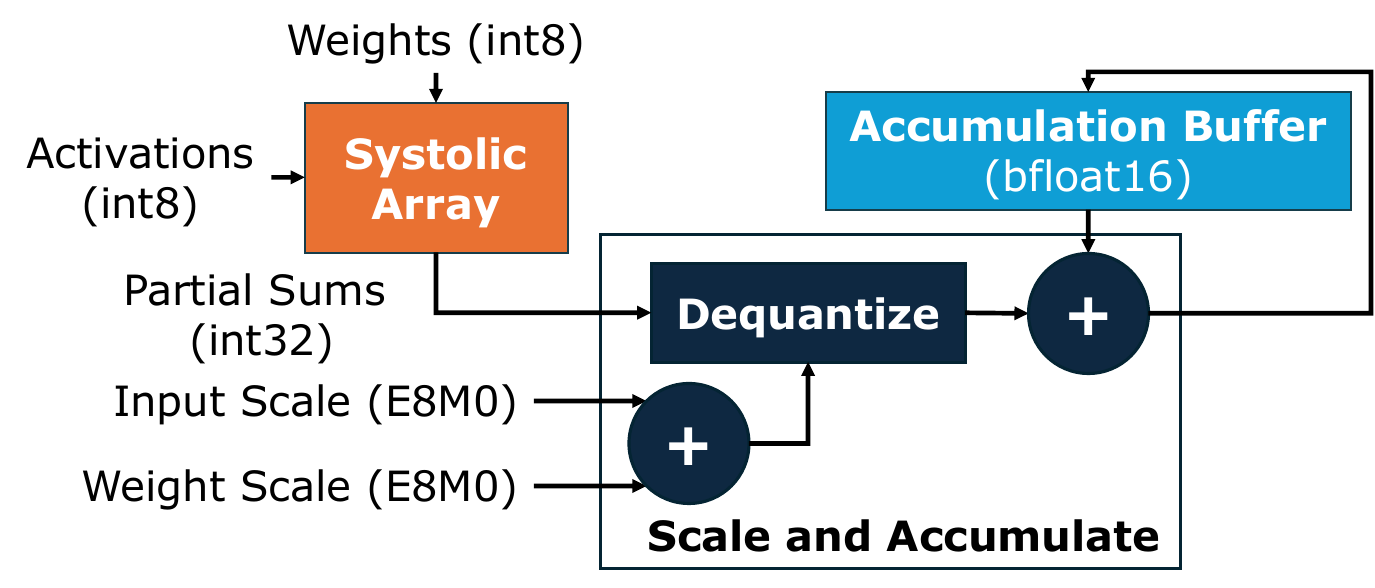}
    \caption{Voyager can generate accelerators with microscaling by introducing dequantization logic in the matrix unit.}
    \label{fig:microscaling-hw}
\end{figure}

Voyager can also generate accelerators that support \textbf{microscaling}~\cite{microscaling}, a quantization scheme that applies block-wise scaling along the reduction dimension to improve accuracy in low-precision GEMM operations. Supporting this scheme requires specialized hardware to both efficiently perform the GEMM operation and to generate scaling factors per output block. As shown in \autoref{fig:microscaling-hw}, Voyager inserts additional dequantization logic within the matrix unit. Inputs and weights are stored in a low-precision integer format, such as INT8, and each have an associated scale factor represented in E8M0 format. The systolic array accumulates in a higher precision integer format, such as INT32, and dequantizes the partial sum to a floating-point format, such as BF16, using the product of the input and weight scales. This scaled partial sum is then accumulated with a previous partial sum and stored in the accumulation buffer. 

  \begin{figure}
    \centering
    %\begin{subfigure}[b]{1.0\columnwidth}
        %\centering
        \begin{lstlisting}[language=C++, frame=none, numbers=none]
// Sub-optimal C++ coding style for HLS
for(int ic = 0; ic < IC_loop_bound; ic++){
 for(int oy = 0; oy < OY_loop_bound; oy++){
  for(int ox = 0; ox < OX_loop_bound; ox++){
   int address = oy * OX_loop_bound * IC_loop_bound
               + ox * IC_loop_bound + ic;
   ...
  }
 }
}
// Coding style for improved results     
int IC_last = IC_loop_bound - 1;
int OY_last = OY_loop_bound - 1;
int OX_last = OX_loop_bound - 1;
int OY_stride = OX_loop_bound * IC_loop_bound;
for(int ic = 0;; ic++){
 for(int oy = 0;; oy++){
  for(int ox = 0;; ox++){
   int address = oy * OY_stride
               + ox * IC_loop_bound + ic;
   ...
   if(ox == OX_last) { break; }}
  if(oy == OY_last) { break; }}
 if(ic == IC_last) { break; }}
        \end{lstlisting}
        %\caption{Coding style for improved results.}
        %\label{subfig:optimal_code}
    %\end{subfigure}
    \caption{Coding style has a large impact on HLS-generated RTL. We restructure loops from a standard nested loop (top) to one with constant precomputation and breaks (bottom).}
    \label{fig:optimizations}
\end{figure}

\subsection{Scheduling: Loop Transformations}
While previous axes define what hardware is generated, the scheduling axis determines how computation is mapped to that hardware. Voyager generates accelerators with runtime-configurable loop counters and address generators that control data movement and memory access patterns. These controllers allow the accelerator to support a wide range of loop ordering and tiling strategies. Scheduling directly impacts data reuse, which in turn affects energy efficiency and latency. The compiler’s strategy for schedule selection and lowering is described in Section~\ref{section:compiler_scheduling}.

\subsection{HLS C++ Coding Style Optimizations}

The C++ coding style used in the generator has a significant impact on the quality of the RTL generated from HLS, affecting area, timing, and HLS tool's runtime. Voyager's address generators are implemented using nested for loops, with address computation logic in the loop body. To improve HLS results, we perform two optimizations shown in \autoref{fig:optimizations}:
\textbf{(1) Pre-computation of constants}: Any values that can be computed outside the loop nests, such as loop bounds or loop strides, are moved out of the body. This reduces the dependency chain of operations within the pipeline, improving timing. Additionally, this reduces the number of arithmetic operations that the HLS tool must analyze while scheduling.
\textbf{(2) Flattened loop control}: We restructure the loops as infinite loops with explicit break conditions in the body. This avoids the need for a final extra iteration introduced by the HLS tool when evaluating loop bounds in the loop header. This speeds up the inner loop which has a significant impact, since there are many levels of nested loops.

\section{Voyager's Compiler}
\begin{figure}[t!]
    \centering
    \includegraphics[width=1\linewidth]{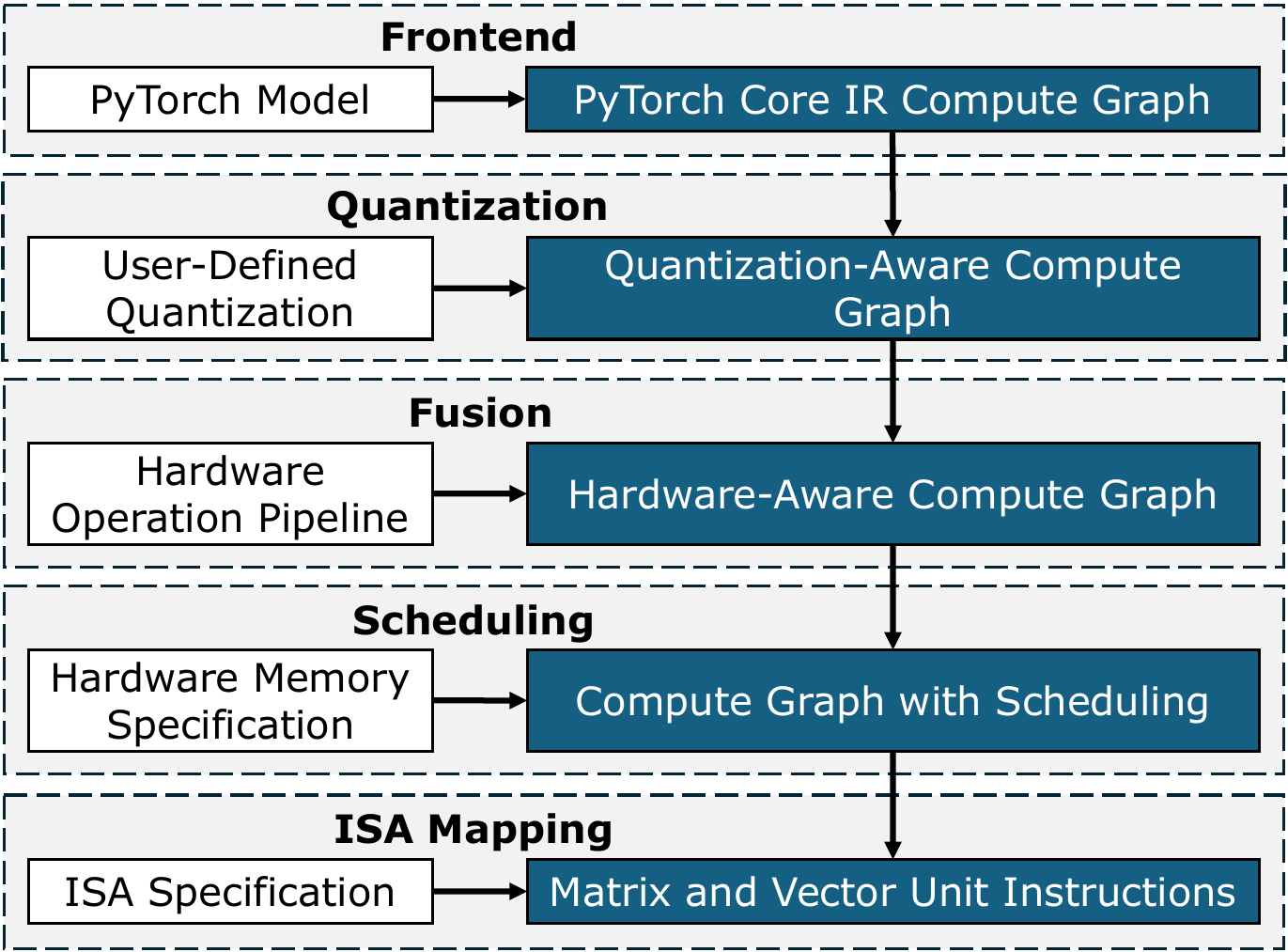}
    \caption{The Voyager compiler lowers PyTorch models to hardware instructions through quantization, hardware-specific optimizations such as fusion and tiling, and generation of matrix/vector instructions for the accelerator ISA.}
    \label{fig:compiler_flow}
\end{figure}

To efficiently deploy models onto Voyager-generated accelerators, we co-designed a compiler with the hardware template. Like many DNN accelerators, Voyager uses a domain-specific instruction set, requiring a custom compiler to transform high-level models into optimized hardware instructions. The compiler consumes PyTorch models and lowers them through stages of quantization, hardware-aware optimizations, and scheduling, as shown in \autoref{fig:compiler_flow}.

To provide a visual understanding of Voyager's software compilation flow, we show how a simple example PyTorch model, with a convolution layer followed by a ReLU, gets lowered to the hardware accelerator's ISA.

\subsection{Frontend}

\begin{figure}[t!]
\centering

% Top: code snippet
\adjustbox{valign=t}{
\begin{minipage}{\linewidth}
\centering \textbf{Example PyTorch Model}
\begin{lstlisting}[language=Python, numbers=none, basicstyle=\ttfamily\small, frame=none]
class Example(nn.Module):
  def __init__(self):
    super(Example, self).__init__()
    self.conv = nn.Conv2d(
      in_channels=32,
      out_channels=64,
      kernel_size=3,
      stride=1)
    self.relu = nn.ReLU()
  def forward(self, x):
    x = self.conv(x)
    x = self.relu(x)
    return x
\end{lstlisting}
\end{minipage}
}

% Bottom: image
\adjustbox{valign=t}{
\begin{minipage}{\linewidth}
\centering
\textbf{Static Compute Graph}

\includegraphics[width=0.5\linewidth]{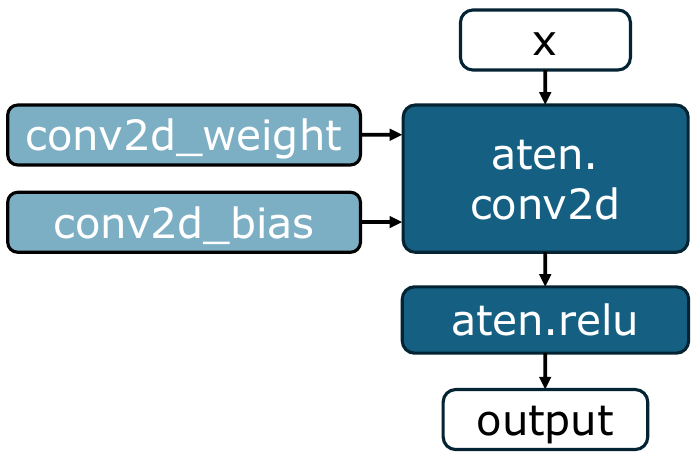}
\end{minipage}
}
\caption{The frontend extracts a static compute graph from a PyTorch model using PyTorch 2 Export. The original model (top) is lowered to an intermediate representation consisting of ATen operators (bottom), the input to later compiler stages.}
\label{fig:model-and-frontend}
\end{figure}

Our compiler's frontend converts PyTorch models into an intermediate representation (IR) that bridges the gap between the high-level model and the accelerator’s instruction set architecture (ISA). We use PyTorch 2 Export (PT2E) \cite{pytorch2} to extract a static compute graph, as shown in \autoref{fig:model-and-frontend}. We adopt PyTorch as the frontend due to its widespread use and ecosystem, including torchvision~\cite{torchvision} and Hugging Face Transformers~\cite{huggingface-transformers}. Users can import existing models into the Voyager toolchain with minimal modification.

\subsection{Quantization}

\begin{figure}
\centering

% Top: code snippet
\adjustbox{valign=t}{
\begin{minipage}{\linewidth}
\centering \textbf{Quantization Specification}
\begin{lstlisting}[language=Python, numbers=none, basicstyle=\ttfamily\small, frame=none]
quantizer = get_default_quantizer(
  inputs="int8,qs=per_tensor",
  weight="int8,qs=per_tensor",
  bias="int32",
)
model = pt2e(model, quantizer)
\end{lstlisting}
\end{minipage}
}

% Bottom: image
\adjustbox{valign=t}{
\begin{minipage}{\linewidth}
\centering
\textbf{Quantization-Aware Compute Graph}

\includegraphics[width=1\linewidth]{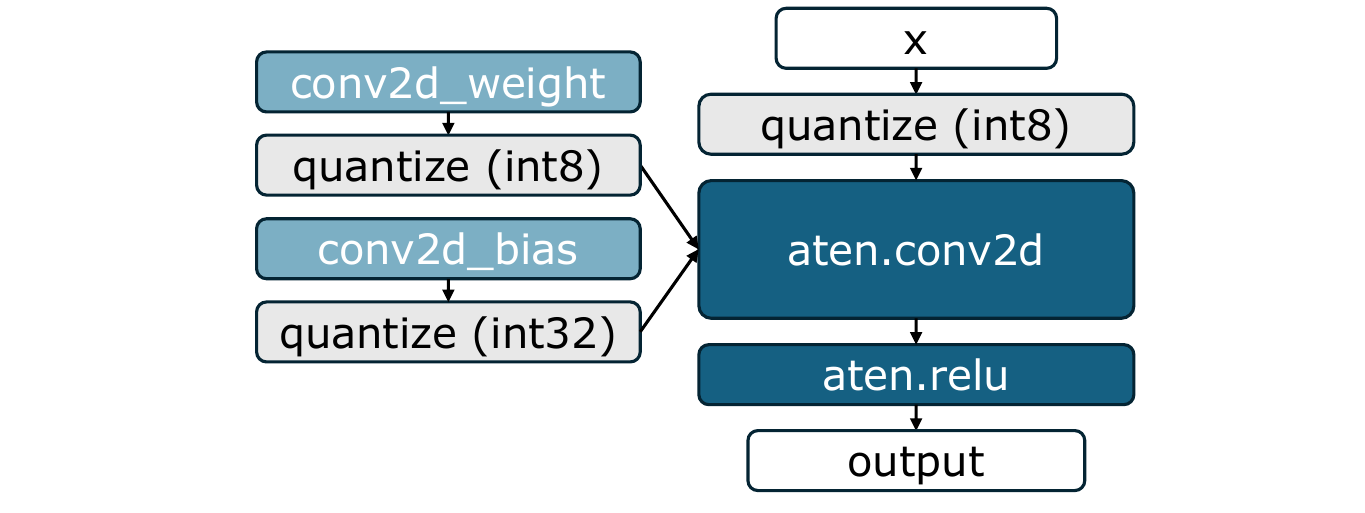}
\end{minipage}
}
\caption{Voyager inserts quantization operators into the compute graph to emulate quantized execution. The user specifies a quantization configuration \textbf{(top)}, which is applied to the model and results in a quantization-aware compute graph \textbf{(bottom)}.}
\label{fig:compiler_quantization}
\end{figure}

While most ML frameworks natively support quantization, including post-training quantization and quantization-aware training, they support only a limited set of datatypes and quantization schemes. They do not offer built-in support for state-of-the-art quantization techniques. To address this, our compiler incorporates a custom quantization framework, where users have fine-grained control over quantization behavior for weights, activations, and biases, as shown in \autoref{fig:compiler_quantization}. The quantizer annotates graph nodes with quantization intent and inserts fake quantization operators to emulate quantized behavior in the floating-point model.

\textit{\textbf{Data Types.}} Voyager extends datatype support to arbitrary-bitwidth integers, floating-point, posit, and NormalFloat. Users can define custom types through a fake quantization function that maps BFloat16 values to the custom type. 

% Since many of these data types are not supported on commercial hardware (CPU, GPU, or NPU), executing quantized networks, particularly large language models, can be prohibitively slow. To address this, Voyager accelerates quantization by creating a dictionary that maps BFloat16 values into their quantized counterparts.

\textit{\textbf{Quantization Schemes.}} We support a wide range of quantization schemes. While PyTorch provides per-tensor and per-channel scaling, which are commonly used for int8 and FP8 quantization, these methods fail to maintain accuracy as precision drops below 8 bits. Microscaling, also known as block-wise quantization, has shown notable accuracy improvements for sub-8-bit quantization. Smaller block sizes generally offer higher accuracy but at the cost of increased bits-per-parameter and computation. To our knowledge, ours is the first compiler to support microscaling with fully configurable block sizes and data types.

\textit{\textbf{Quantization Granularity.}} We provide fine-grained quantization control, allowing different datatypes and quantization schemes per layer based on hardware constraints and performance goals. This is essential for mixed-precision quantization, where less sensitive layers can use lower precision to reduce model size. 

% Users can customize data types and quantization schemes for individual layers by specifying criteria such as operator type, module type, module name, or the order of an operator within a module. These filtering options allow precise targeting of individual operations or modules within a neural network, enabling flexible, efficient quantization.

\begin{figure}
    \centering
    \includegraphics[width=1\linewidth]{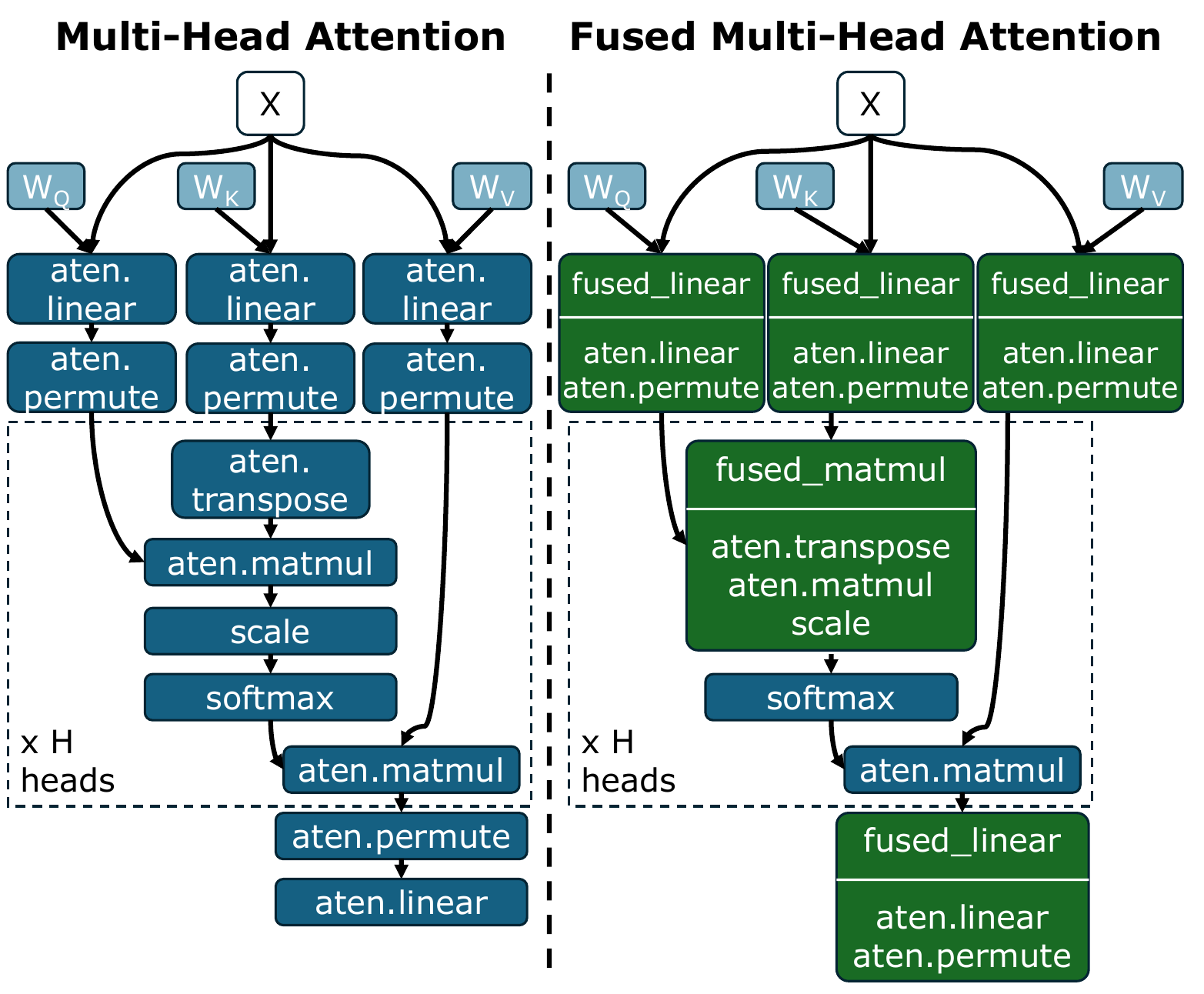}
    \caption{Reshape and matmul fusion in self-attention.}
    \label{fig:attention_fusion}
\end{figure}

\subsection{Fusion}
Voyager performs three types of fusion: quantization fusion, operation fusion, and reshape fusion.

\begin{figure}
    \centering
    \textbf{Fused Compute Graph}
    \includegraphics[width=1\linewidth]{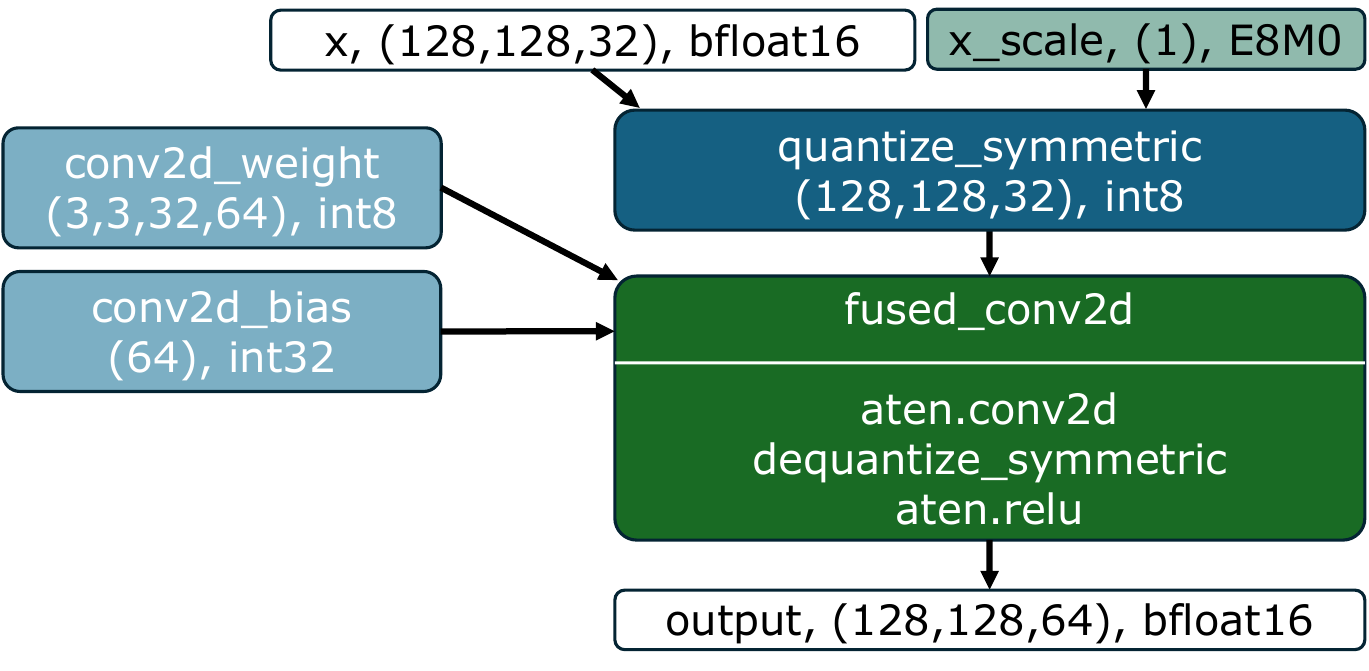}
    \caption{The compiler performs fusion of quantization and operation nodes. In this example, it replaces convolution-dequantization-ReLU with a single fused operation.}
    \label{fig:compiler_fusion}
\end{figure}

\textit{\textbf{Quantization Fusion.}}
After quantization, our compiler replaces fake quantization modules with hardware-executable quantize and dequantize operators. Quantize scales a floating-point tensor and then casts it to the target data type, while dequantize converts a quantized tensor back to floating-point by dividing it by the corresponding scale. Scale factors are computed from tensor statistics observed during calibration runs over sample data. Voyager supports both power-of-two scaling for better hardware efficiency and general floating-point scaling for higher accuracy. When direct quantization is used (e.g., scale factor of 1), redundant dequantize operators are removed. To enable integer-only GEMM, dequantization is deferred until after GEMMs, keeping  computation in integer format for improved energy efficiency.

\textit{\textbf{Operation Fusion.}} 
The compiler combines multiple smaller operations into a single kernel, avoiding intermediate writes to memory and increasing hardware utilization. We support fusing matrix operations with subsequent element-wise vector operations, as well as across multiple vector operations. Fusion is guided by the vector pipeline structure: the compiler replaces sequential patterns that match the supported hardware pipeline and replaces them with fused instructions. This transformation is illustrated in \autoref{fig:compiler_fusion}, where a sequence of convolution, dequantization, and ReLU activation is replaced with a single fused operator.

\textit{\textbf{Reshape Fusion.}}
The compiler also fuses reshape and dequantize operations with their consumers. The vector unit's address generators can directly perform reshaping operations such as transpositions and axis permutations during memory access, eliminating intermediate tensors and reducing both latency and memory overhead. This is especially beneficial for transformers, where multi-head attention involves several reshapes: query, key, and value projections are permuted to separate the head dimension, the key is transposed, and outputs of the attention heads are recombined by another permutation. As shown in \autoref{fig:attention_fusion}, the compiler fuses linear projections with permutation and transpose, producing a self-attention graph with fewer memory accesses and lower latency. 

\subsection{Accelerator IR Generation}

\begin{figure}
    \centering
    \begin{lstlisting}[language=Python, numbers=none, basicstyle=\ttfamily\small, frame=none]
      name: "conv2d_fused"
      op_list: [{
       name: "conv2d"
       target: "conv2d"
       input:
        node: "quantize_symmetric"
        shape: [128,128,32]
        dtype: "int8"
        address: 18688
       weight:
        node: "conv2d_weight"
        shape: [3,3,32,64]
        dtype: "int8"
        address: 0
       bias: 
        node: "conv2d_bias"
        shape: [64]
        dtype: "int32"
        address: 18432
      },
       {
        name: "dequantize_default"
        target: "dequantize"
        dtype: "int24"
        input:
         node: "conv2d"
         shape: [128,128,64]
         dtype: "int32"
        scale: 
         node: "conv2d_scale"
         shape: [1]
         dtype: "bfloat16"
       },
       {
        name: "relu"
        target: "relu"
        input: 
         node: "dequantize_default"
         shape: [128,128,64]
         dtype: "bfloat16"
      }]
      output:
       node: "conv2d_fused"
       shape: [128,128,64]
       dtype: "bfloat16"
       address: 542976
    \end{lstlisting}
    \caption{The compiler generates an intermediate representation (IR) consisting of operations, tensor metadata (shape, datatype), and memory addresses.}
    \label{fig:compiler_ir}
\end{figure}

To enable downstream optimizations and hardware mapping, the Voyager compiler generates a hardware-oriented IR tailored for Voyager-generated accelerators. As illustrated in \autoref{fig:compiler_ir},
The IR captures both the high-level operations and low-level details, such as datatypes, tensor shapes, and memory addresses. 
The core unit is a \texttt{Tensor} object, mirroring a PyTorch tensor but enriched with hardware-relevant metadata, such as shape, datatype, memory address, and an optional permutation field to support implicit reshaping (e.g., transpositions).

PyTorch operations are categorized into four main groups: (1) \textbf{matrix operations} including convolution and general matrix multiplication (GEMM), (2) \textbf{elementwise operations} such as ReLU, GeLU, and residual additions, (3) \textbf{complex reductions} like LayerNorm and softmax, and (4) \textbf{reshape operations} like permute, view, and transpose. Each graph node maps to an IR block of type (1), (3), or (4). If a node represents a fused operation, its element-wise functions (category 2) are appended as a post-processing list in the same IR block. For example, a convolution followed by residual addition and ReLU is represented as a matrix operation with fused element-wise transformations. 

\subsection{Scheduling}
\label{section:compiler_scheduling}

\begin{figure}
    \centering
    \begin{lstlisting}[language=Python, numbers=none, basicstyle=\ttfamily\small, frame=none]
# L2 level
for oy1 in range(16):
 for ox1 in range(4):
  for oc2 in range(2):
  # Accumulation buffer level
   for ic1 in range(2):
    for oc1 in range(2):
     for fx in range(3):
      for fy in range(3):
       for oy0 in range(8):
        for ox0 in range(32):
         # Systolic array level
         parallel_for ic0 in range(16):                          
         parallel_for oc0 in range(16):
         
          # Compute shape
          input_height = 16*8
          input_width = 4*32
          
          # Compute indices
          oy = oy1 * 8 + oy0
          ox = ox1 * 32 + ox0
          oc = oc2 * 2 * 16 + oc1 * 16 + oc0
          ic = ic1 * 16 + ic0
          
          # Adjust for padding
          in_y = oy - 1
          in_x = ox - 1
          
          # Add bias
          if ic == 0 and fx == 0 and fy == 0:
           output[oy,ox,oc] = bias[oc]
           
          # Perform MAC
          if (0 <= in_y < input_height) and 
             (0 <= in_x < input_width):
           output[oy,ox,oc] += input[in_y,in_x,ic] 
                             * weight[fy,fx,ic,oc]
                             
          # Apply ReLU activation
          if ic==31 and fy==2 and fx==2:
           output[oy,ox,oc] = max(output[oy,ox,oc],0)
    \end{lstlisting}
    \caption{An example loop schedule generated using an extended version of Interstellar.}
    \label{fig:example-schedule}
\end{figure}

Mapping DNNs onto hardware involves making scheduling decisions such as loop tiling, loop ordering, and loop fusion. The space of possible schedules is large, with each producing distinct data reuse patterns across memory levels. Optimizing one type of data reuse reduces opportunities for another. Voyager addresses this by extending Interstellar~\cite{interstellar} to efficiently schedule DNN layers onto the generated hardware.

Interstellar uses a coarse-grained analytical cost model that estimates memory energy by counting accesses at each level of the memory hierarchy and applying per-access energy weights from a hardware-specific lookup table. A heuristic search then explores the schedule space efficiently.

We extend Interstellar in two ways: (1) we constrain the schedule space to only the schedules feasible on Voyager-generated accelerators, and (2) add a performance model that accounts for runtime based on for hardware characteristics. 

Voyager's performance model breaks down execution time into two components: \textbf{(1) L1 tile compute time}: This is based on the tile size and the number of compute elements in the systolic array. Before computation, weights must be loaded into the systolic array, requiring one cycle per row in the array. To hide this latency, the weights within PEs are double-buffered, so weight loading for the next tile is overlapped with computation of the current tile. Thus, compute time per tile is the maximum of (a) the time to perform the computation, set by the tile size and PE latency, and (b) weight loading time. The total latency for an L1 tile equals this per-tile latency multiplied by the number of systolic array tiles. \textbf{(2) L1 tile memory access time}: This is how long it takes to transfer tiled input and weight tensors into the L1 buffers, determined by the memory bandwidths.

The total runtime is the maximum of the L1 compute and memory access time, multiplied by the number of tiles in L2 memory. If the design lacks a double-buffered accumulation buffer, the model also accounts for vector unit backpressure by including its compute time in the maximum. For deeper memory hierarchies, this process is repeated at each level. By incorporating these hardware-specific effects, our performance model provides fast and accurate runtime predictions.

An example of a generated schedule is shown in \autoref{fig:example-schedule}, which illustrates a multi-level loop nest for a convolution.

\subsection{Accelerator Instruction Generation}

\begin{figure}
    \centering
    \begin{lstlisting}[language=C++, numbers=none, basicstyle=\ttfamily\small, frame=none]
        struct MatrixUnitInst {
          uint64_t input_base_address = 18688;
          uint64_t weight_base_address = 0;       
          uint64_t bias_base_address = 18432;
        
          uint16_t loops[2][6] = 
          {
            {16,4,2,1}, // L2 level
            {2,2,3,3,8,32} // accum buffer level
          };
          uint3_t x_loop_index[2] = {1,5};
          uint3_t y_loop_index[2] = {0,4};
          uint3_t reduction_loop_index[2] = {3,0};
          uint3_t output_channel_loop_index[2] = 
            {2,1};
          uint3_t fx_index = 2;
          uint3_t fy_index = 3;
          ...
        };
    \end{lstlisting}
    \caption{A subset of a matrix unit instruction used to configure memory addresses, loop bounds, and loop orderings.}
    \label{lst:mu-inst}
\end{figure}

Using the accelerator IR and the schedule from Interstellar, the compiler maps high-level operations onto Voyager hardware instructions. Matrix unit instructions encode loop configuration fields, such as bounds, orderings, and strides. Interstellar's loop nests map  directly to this structure, allowing schedules to be translated into address generator configurations with minimal transformation. Vector unit instructions also include address generator configurations, but extend to specify the vector pipeline configuration. This includes selection of functional units for each stage, as well as settings for the accumulation and reduction engines, and quantization/dequantization. Together, these instructions enable the compiler to efficiently map diverse operations, including fused patterns, onto the accelerator.

\autoref{lst:mu-inst} shows a subset of the matrix unit instruction. The \texttt{base address} fields specify the memory addresses for tensors in main memory. The \texttt{loops} field encodes loop bounds at two levels of hierarchy, as specified by the schedule generated by Interstellar. The \texttt{*\_loop\_index} fields map each dimension to its respective index in the \texttt{loops} field.

\section{Results}
\begin{comment}
\begin{table}
    \centering
    \begin{tabular}{c|c|c}
    \hline
    \textbf{Coding Style} & \textbf{Frequency (MHz)} & \textbf{Area (mm\textsuperscript{2}) } \\
    \hline
         Unoptimized & 400 & 1.16 \\
         Optimized & 400 & 1.18 \\
         Optimized & 1000 & 1.41 \\
    \hline
    \end{tabular}
    \caption{Impact of HLS coding style on frequency and area.}
    \label{tab:coding-style}
\end{table}
\end{comment}
\begin{table}
\centering
\begin{tabular}{ll|c|c}
    \hline
    &  & \textbf{Runtime} & \textbf{Speedup} \\
    \hline
    \multirow{4}{*}{\rotatebox[origin=c]{90}{Fusion}} & Baseline & 2.67 ms & - \\
    & \quad + Permute \& Transpose & 2.65 ms & 0.75\% \\
    & \quad + Residual & 2.61 ms & 2.30\% \\
    & \quad + NoNorm & 2.37 ms & 12.66\% \\
    \hline
    \multirow{3}{*}{\rotatebox[origin=c]{90}{Model}} & Energy-Only & 2.44 ms & - \\
    & Coarse-Grained Perf. Model & 2.44 ms & - \\
    & Fine-Grained Perf. Model & 2.37 ms & 2.86\% \\
    \hline
    & Without Double Buffer & 2.37 ms & - \\
    & With Double Buffer & 2.29 ms & 3.37\% \\
    \hline
\end{tabular}
\caption{Runtime improvement from operator fusion, fine-grained performance modeling and double buffering.}
\label{tab:fusion_results}
\end{table}

\begin{comment}
\begin{table}
\centering
\begin{tabular}{l|c|c}
\hline
\textbf{Fusion} & \textbf{Runtime (ms)} & \textbf{Speedup} \\
\hline
Baseline & 2.67 & -- \\
\quad + Permute \& Transpose & 2.65 & 0.75\% \\
\quad + Residual & 2.61 & 2.30\% \\
\quad + NoNorm & 2.37 & 12.66\% \\
\hline
\end{tabular}
\caption{Runtime improvement from operator fusion.}
\label{tab:fusion_results}
\end{table}

\begin{table}
    \centering
    \begin{tabular}{l|c}
    \hline
    \textbf{Schedule Selection Strategy} & \textbf{Runtime (ms)} \\
    \hline
    Energy-Only Scheduling & 2.44 \\
    Coarse-Grained Performance Model & 2.44 \\
    Full-Fidelity Performance Model & 2.37 \\
    \hline
    \end{tabular}
\caption{Impact of performance modeling fidelity.}
    \label{tab:perf_model_results}
\end{table}

\begin{table}
    \centering
    \begin{tabular}{l|c}
    \hline
    \textbf{Configuration} & \textbf{Runtime (ms)} \\
    \hline
    Without Double Buffer & 2.37 \\
    With Double Buffer & 2.29 \\
    \hline
    \end{tabular}
    \caption{Effect of double buffering of accumulation buffer.}
    \label{tab:double_buffering_results}
\end{table}
\end{comment}

\begin{table*}
\centering
\begin{threeparttable}
\begin{tabular}{l|l|l|l|llllll}
\hline
\multicolumn{1}{c|}{\multirow{2}{*}{\textbf{Type}}} & \multicolumn{1}{c|}{\multirow{2}{*}{\textbf{Model}}} & \multirow{2}{*}{\textbf{Task (Dataset)}} & \multirow{2}{*}{\textbf{Size}} & \multicolumn{6}{c}{\textbf{Accuracy}} \\ \cline{5-10}
\multicolumn{1}{c|}{} & \multicolumn{1}{c|}{} &  &  & FP32 & BF16 & E4M3 & Posit8 & INT8 & MXINT8 \\ \hline
\multirow{4}{*}{Vision} 
& ResNet-18 & Image Classification (ImageNet) & 11.7M & 71.5\% & 70.0\% & 69.2\% & 69.1\% & 71.5\% & 70.7\% \\
& ResNet-50 & Image Classification (ImageNet) & 25.6M & 80.4\% & 81.1\% & 78.8\% & 79.4\% & 78.7\% & 79.8\% \\
& ViT (Base) & Image Classification (ImageNet) & 86M & 84.1\% & 84.3\% & 83.8\% & 84.0\% & 75.4\% & 84.0\% \\
& MobileNetV2 & Image Classification (ImageNet) & 3.5M & 71.1\% & 70.1\% & 62.5\% & 63.8\% & 66.0\% & 71.3\% \\ \hline
\multirow{2}{*}{NLP} 
& MobileBERT-TINY & Sentiment Classification (SST-2) & 15M & 90.8\% & 90.7\% & 90.6\% & 90.4\% & 90.4\% & 91.1\% \\
& BERT (Base) & Sentiment Classification (SST-2) & 110M & 93.2\% & 93.0\% & 93.1\% & 92.8\% & 92.4\% & 93.1\% \\
\hline
\end{tabular}
\end{threeparttable}
\caption{Accuracy of vision and NLP models across different data types. NLP models are run with a sequence length of 128. }
\label{tab:accuracy}
\end{table*}

\begin{table*}[]
\centering
\begin{tabular}{c|c|c|c|c|c|c|c|c}
\hline
\multirow{3}{*}{\textbf{Model}} & \multirow{3}{*}{\textbf{Metric}} & \multicolumn{7}{c}{\textbf{Array Size and Total SRAM Capacity}} \\
\cline{3-9}
& & 8x8 & 8x16 & 16x16 & 16x32 & 32x32 & 32x64 & 64x64 \\
& & (48 KB) & (80 KB) & (96 KB) & (160 KB) & (192 KB) & (320 KB) & (384 KB) \\
\hline
\multirow{2}{*}{ResNet-18} & Runtime (cycles) & 31.76M & 15.88M & 7.90M & 3.96M & 2.07M & 1.04M & 660K \\
& MAC Utilization (\%) & 90.9 & 90.9 & 93.0 & 92.9 & 91.9 & 91.4 & 76.8 \\
\hline
\multirow{2}{*}{ResNet-50} & Runtime (cycles) & 67.70M & 33.86M & 17.00M & 8.51M & 4.40M & 2.21M & 1.30M \\
& MAC Utilization (\%) & 95.4 & 95.4 & 96.2 & 96.1 & 94.9 & 94.5 & 82.1 \\
\hline
\multirow{2}{*}{ViT (Base)} & Runtime (cycles) & 326.19M & 163.14M & 84.37M & 42.24M & 22.74M & 12.44M & 7.53M \\
& MAC Utilization (\%) & 99.4 & 99.4 & 99.3 & 99.2 & 98.0 & 96.5 & 88.0  \\
\hline
\multirow{2}{*}{MobileNetV2} & Runtime (cycles) & 7.373M & 3.760M & 2.010M & 1.099M & 659K & 534K & 407K \\
& MAC Utilization (\%) & 86.7 & 84.9 & 83.7 & 79.3 & 78.0 & 58.6 & 52.5  \\
\hline
MobileBERT-TINY & Runtime (cycles) & 27.91M & 14.00M & 7.71M & 3.91M & 2.37M & 1.28M & 916K \\
 Sequence Length = 128 & MAC Utilization (\%) & 98.3 & 98.0 & 95.1 & 93.8 & 86.2 & 82.8 & 67.3 \\
\hline
BERT (Base) & Runtime (cycles) & 179.17M & 89.63M & 46.07M & 23.09M & 12.24M & 6.48M & 3.80M \\
 Sequence Length = 128 & MAC Utilization (\%) & 99.8 & 99.8 & 99.4 & 99.2 & 97.9 & 96.2 & 90.7 \\
\hline
LLaMA 3.2 1B Prefill & Runtime (cycles) & -- & -- & -- & -- & 688.0M & -- & -- \\
Sequence Length = 512& MAC Utilization (\%) & -- & -- & -- & -- & 99.8 & -- & -- \\
\hline
\end{tabular}
\caption{Runtime and utilization across E4M3-based designs at different scales. }
\label{tab:scaling}
\end{table*}

We evaluate Voyager across multiple design points using DNNs models from different application domains. Our evaluation flow includes high-level synthesis using Catapult HLS to generate RTL Verilog, cycle-accurate simulations using Synopsys VCS, and logic synthesis using Synopsys Design Compiler. MAC utilization is defined as the ideal cycles divided by the achieved cycles, where the ideal cycles is the total number of MAC operations in a layer divided by the number of hardware MAC units. In Section~\ref{section:voyager_optimizations}, we analyze the impact of key design decisions within Voyager. Section~\ref{section:dse} shows how we can use Voyager for design space exploration. Finally, Sections~\ref{section:comparison_gen} and ~\ref{section:comparison_hand} compare Voyager against state-of-the-art hardware generators and hand-optimized designs.

\subsection{Optimizations within Voyager}
\label{section:voyager_optimizations}

We evaluate the impact of optimizations in Voyager using a 32$\times$32 E4M3 design in TSMC 16 nm on MobileBERT-TINY.

\textit{\textbf{Coding Style Optimizations.}} With the unoptimized coding style, the design is limited to a maximum frequency of 400 MHz, with an area of 1.16 mm$^2$. In contrast, the optimized coding style substantially improves timing closure, enabling a maximum frequency of up to 1 GHz, with an area of 1.41 mm$^2$. When normalized to the same frequency (400 MHz), the optimized design has an area of 1.18 mm$^2$, only a slight area increase of 1.7\% versus the unoptimized version.

\textit{\textbf{Operator Fusion.}} Voyager can fuse operators, which reduces memory accesses and improves latency. \autoref{tab:fusion_results} shows the impact of various degrees of fusion on latency, with a final runtime improvement of 12.66\%.

\textit{\textbf{Scheduling.}} %Voyager uses a fine-grained performance model to guide scheduling decisions. 
%Unlike prior approaches that assume ideal compute utilization~\cite{interstellar} or rely on coarse-grained estimates~\cite{timeloop}, it models hardware characteristics to generate more performant schedules that better exploit the hardware. 
\autoref{tab:fusion_results} compares runtime using three scheduling strategies: an energy-only heuristic, a coarse-grained performance model, and our fine-grained performance model, which achieves up to 2.9\% lower runtime, showing the value of accurate performance modeling in schedule selection.

\textit{\textbf{Double Buffering.}} The tight coupling between the matrix and  vector units can result in slowdowns in situations where the vector unit is bottlenecked by memory accesses. To mitigate this, Voyager supports an optional double-buffered accumulation buffer, decoupling its output from the vector unit's input, leading to a 3\% reduction in runtime (\autoref{tab:fusion_results}).

\subsection{Design Space Exploration with Voyager}

\label{section:dse}
To show Voyager's design space exploration capability we present how we can use it to explore (1) different datatypes and quantization schemes (resource type) and (2) accelerator sizes (resource allocation) on various DNNs. For these studies, we use the optimal schedules from our compiler.

\textit{\textbf{Exploring Datatypes and Quantization Schemes.}} Using the bit-accurate C++ model, Voyager allows running inference on entire datasets to evaluate hardware-level accuracy on different DNNs for different datatypes and quantizations. \autoref{tab:accuracy} shows the metrics for vision and NLP models quantized to different types (FP32, BF16, E4M3, Posit8, INT8 and MXINT8). We use pre-trained models from torchvision~\cite{torchvision}, timm~\cite{timm}, and transformers~\cite{huggingface-transformers} libraries. These results show that microscaling (MXINT8) enables high-accuracy integer computation, achieving accuracy on par with floating-point formats. Voyager makes it possible to evaluate this across diverse models. 
However, using microscaling can come at an area cost. Voyager allows us to evaluate this as well. \autoref{tab:area-breakdown} reports the area of Voyager-generated accelerators in TSMC 16 nm at 1 GHz clock frequency. 
%Floating-point formats require almost double the systolic array area of INT8, while INT8 consumes more area for memory to support wider accumulations. 
MXINT8 requires 43\% more area for addressing and control than INT8 (for a 32$\times$32 design) due to the added support for input and weight scale factors. However, its overall area is still 5.64\% lower than INT8 due to a smaller accumulation buffer and vector unit, and 11.6\% lower than E4M3 due to a smaller systolic array.

\begin{table}[t!]
\centering
\begin{tabular}{c|c|ccccc}
\hline
\multirow{3}{*}{\rotatebox[origin=c]{90}{\textbf{Size}}}
 & \multirow{3}{*}{\textbf{Type}}
 & \multicolumn{5}{c}{\textbf{Area (mm$^2$)}} \\
\cline{3-7}
& & \makecell{Sys.\\Array} & SRAM & \makecell{Addr.\\\& Ctrl.} & \makecell{Vector\\Unit} & Total \\
\hline
\multirow{4}{*}{\rotatebox[origin=c]{90}{8x8}} & E4M3 & 0.048 & 0.085 & 0.062 & 0.142 & 0.336 \\
& Posit8 & 0.052 & 0.085 & 0.064 & 0.145 & 0.346 \\
& INT8 & 0.027 & 0.111 & 0.064 & 0.148 & 0.350 \\
& MXINT8 & 0.031 & 0.095 & 0.095 & 0.135 & 0.355 \\
\hline
\multirow{4}{*}{\rotatebox[origin=c]{90}{16x16}} & E4M3 & 0.192 & 0.109 & 0.096 & 0.237 & 0.633 \\
& Posit8 & 0.210 & 0.109 & 0.101 & 0.242 & 0.661 \\
& INT8 & 0.109 & 0.211 & 0.098 & 0.250 & 0.667 \\
& MXINT8 & 0.123 & 0.123 & 0.161 & 0.224 & 0.632 \\
\hline
\multirow{4}{*}{\rotatebox[origin=c]{90}{32x32}} & E4M3 & 0.769 & 0.217 & 0.199 & 0.442 & 1.627 \\
& Posit8 & 0.846 & 0.217 & 0.215 & 0.450 & 1.728 \\
& INT8 & 0.440 & 0.422 & 0.198 & 0.465 & 1.525 \\
& MXINT8 & 0.501 & 0.243 & 0.283 & 0.412 & 1.438 \\
\hline
\multirow{4}{*}{\rotatebox[origin=c]{90}{64x64}} & E4M3 & 3.104 & 0.434 & 0.453 & 0.825 & 4.817 \\
& Posit8 & 3.399 & 0.434 & 0.506 & 0.836 & 5.175 \\
& INT8 & 1.783 & 0.844 & 0.432 & 0.869 & 3.928 \\
& MXINT8 & 2.021 & 0.484 & 0.565 & 0.752 & 3.821 \\
\hline
\end{tabular}
\caption{Post-synthesis area of different accelerator configurations in 16 nm targeting a clock frequency of 1 GHz.}
\label{tab:area-breakdown}
\end{table}

%\subsubsection{Runtime Results}

% \begin{table}[t]
% \centering
% \begin{threeparttable}
% \begin{tabular}{l|r|r}
% \hline
% \textbf{Model} & \textbf{Runtime (ms)} & \textbf{MAC Util. (\%)} \\ 
% \hline
% ResNet-18 & 2.07 & 91.9 \\
% ResNet-50 & 4.40 & 94.9 \\
% ViT (Base) & 22.70  & 98.0 \\
% MobileNetV2 & 0.66 & 78.0 \\
% BERT (Base)\tnote{a} & 12.24 & 97.9 \\
% MobileBERT-TINY\tnote{a} &  2.37 & 86.2 \\
% LLaMA 3.2 1B \tnote{b} & 688.00 & 99.8  \\
% \hline
% \end{tabular}
% \begin{tablenotes}
%     \item[a] Sequence length of 128.
%     \item[b] Prefill phase, sequence length of 512.
% \end{tablenotes}

% \end{threeparttable}
% \caption{Runtime and utilization (util.) of selected vision and NLP models at 1 GHz on a 32x32 E4M3 design.}
% \label{tab:runtime-utilization}
% \end{table}

\textit{\textbf{Exploring Accelerator Sizes.}} Voyager designs scale effectively across systolic array sizes (\autoref{tab:scaling}). Runtimes decrease nearly linearly with array size, maintaining a high hardware utilization. 
%Compute-intensive models like ViT and BERT maintain high utilization above 95\% even at large scales. \autoref{tab:runtime-utilization} summarizes the runtime results for a 32$\times$32 E4M3 design running at 1 GHz. Across models, Voyager achieves a high utilization.

\subsection{Comparison with Prior Generators}
\label{section:comparison_gen}

\autoref{tab:comparison_with_generators} compares Voyager designs against those from two other generators, Gemmini and NVDLA. For a fair comparison, we match the technology node, datatype, number of MACs and frequency to the design points reported in prior work. While Gemmini achieves a 28\% lower area, Voyager is significantly more performant, with a 61\% lower runtime. Compared to NVDLA, Voyager designs are 35.91\% to 56.48\% smaller and achieve 20.84\% to 53.15\% lower runtime.

\begin{table}[t!]
\centering
\begin{tabular}{ccccc}
\hline
\textbf{\# MACs} & \multicolumn{2}{c}{\textbf{Area ($\mu$m$^2$)}} & \multicolumn{2}{c}{\textbf{ResNet-50 Time (ms)}} \\
 &  Gemmini & Voyager & Gemmini & Voyager\\
\hline
256 & 116,000 & 160,314 & 43.58 & 17.00\\
\hline
 &  NVDLA & Voyager & NVDLA & Voyager\\
\midrule
2048 & 3,300,000 & 2,114,852 & 3.72 & 2.21\\

1024 & 1,800,000 & 1,102,659 & 6.54 & 4.40\\

512 & 1,400,000 & 852,504 & 10.75 & 8.51\\

256 & 1,000,000 & 456,468 & 21.74 & 17.00\\

128 & 840,000 & 394,511 & 50.00 & 33.86\\

64 & 550,000 & 239,519 & 136.99 & 67.70\\

32 & 510,000 & 221,943 & 277.78 & 130.13 \\
\hline
\end{tabular}
\caption{Comparison of Voyager with Gemmini in Intel 16 and NVDLA in TSMC 16. All designs use INT8 compute at 1~GHz. Areas are standard cell area (compute array only for Gemmini comparison, full design for NVDLA comparison).
}
\label{tab:comparison_with_generators}
\end{table}

\begin{comment}

\begin{table}
\centering
\begin{tabular}{cccl}
\toprule
\textbf{Generator} & \textbf{\# MACs} & \textbf{Area ($\mu$m$^2$)} & \textbf{Runtime (ms)} \\
\midrule
Gemmini & 256 & 116,000 & 43.58 \\
Voyager & 256 & 160,314 & 17.00 (-61\%)\\
\midrule
NVDLA & 2048 & 3,300,000 & 3.72 \\
Voyager & 2048 & 2,114,852 & 2.21 \\
NVDLA & 1024 & 1,800,000 & 6.54 \\
Voyager & 1024 & 1,102,659 & 4.40 \\
NVDLA & 512 & 1,400,000 & 10.75 \\
Voyager & 512 & 852,504 & 8.51 \\
NVDLA & 256 & 1,000,000 & 21.74 \\
Voyager & 256 & 456,468 & 17.00 \\
NVDLA & 128 & 840,000 & 50.00 \\
Voyager & 128 & 394,511 & 33.86 \\
NVDLA & 64 & 550,000 & 136.99 \\
Voyager & 64 & 239,519 & 67.70 \\
NVDLA & 32 & 510,000 & 277.78 \\
Voyager & 32 & 221,943 & 130.13 \\
\bottomrule
\end{tabular}
\caption{Comparison of Voyager with prior generators. All designs use INT8 compute at 1~GHz. Gemmini comparison results are in Intel 16nm, while NVDLA ones are in TSMC 16nm. Reported areas are standard cell area (compute array only for Gemmini comparison, full design for NVDLA comparison). Runtime is for ResNet-50.}
\label{tab:comparison_with_generators}
\end{table}

\end{comment}
\subsection{Comparison with Hand-Optimized Designs}
\label{section:comparison_hand}
We compare Voyager with two prior hand-optimized accelerators: Simba~\cite{simba_vlsi, simba_hotchips, simba_jssc}, a CNN accelerator, and a transformer accelerator with vector-scaled quantization (VSQ) ~\cite{keller_transformer_jssc}. We match design parameters as much as possible. Because the hand-optimized designs report post-layout results, we scale up the area of Voyager  designs by 1.25$\times$ to account for place-and-route overhead.

To compare with Simba, we generate a design in the same TSMC 16nm process with matching MACs, datatype, and buffer sizes (\autoref{tab:comparison_with_simba_vsq}). While Simba reports results at a higher clock frequency, this is due to a much higher voltage than our characterized libraries, making a direct runtime comparison difficult. Comparing latency in cycles, Voyager greatly outperforms Simba, with a ResNet-50 latency of 4.4 million cycles compared to Simba's 8.3 million. Voyager's utilization is 95\% vs. Simba's 48\%. Even with the lower frequency, the Voyager outperforms Simba by 5\% on absolute runtime.

For the transformer VSQ design, we generate a design in TSMC 7 nm  (due to lack of access to 5 nm) with matching number of MACs, as shown in \autoref{tab:comparison_with_simba_vsq}. Despite the older technology node, Voyager can operate at the same clock frequency. On BERT-Base, Voyager achieves a runtime of 13.4 ms, compared to VSQ's 11.4 ms. However, unlike  Voyager, VSQ does not accelerate the entire model, as it lacks support for reshaping, transposition, residual addition, and layer normalization. In terms of MAC utilization, both designs achieve 98\%. These comparisons show that Voyager can match or outperform hand-optimized designs in performance and utilization, while providing greater design automation.

\begin{table}[t!]
\centering
\begin{tabular}{c|cc|cc}
\toprule
\textbf{Metric} & \textbf{Simba} & \textbf{Voyager} & \textbf{VSQ} & \textbf{Voyager}\\
\midrule
\# MACs & 1024 & 1024 & 1024 & 1024\\
Datatype & INT8 & INT8 & INT4 & INT4\\
Technology & 16nm & 16nm & 5nm & 7nm\\
Frequency (GHz) & 1.797 & 1.0 & 0.909 & 0.909\\
Voltage (V) & 1.1 & 0.8 & 0.67 & 0.75\\
Area (mm\textsuperscript{2}) & 3.1 & 1.8 & 0.153 & 0.57 \\
ResNet-50 Cycles & 8.3M & 4.4M & x & x \\
ResNet-50 ms & 4.64 & 4.40 & x & x \\
BERT-Base Cycles & x & x & 10.4M & 12.2M \\
BERT-Base ms & x & x & 11.4 & 13.4 \\
MAC Utilization & 48\% & 95\% & 98\% & 98\% \\
\bottomrule
\end{tabular}
\caption{Comparison of Voyager with Simba and 4-bit VSQ.}
\label{tab:comparison_with_simba_vsq}
\end{table}

\begin{comment}
\begin{table}
\centering
\begin{tabular}{ccc}
\toprule
\textbf{Metric} & \textbf{Simba~\cite{simba_vlsi}} & \textbf{Voyager} \\
\midrule
\# MACs & 256 & 256 \\
Datatype & INT8 & INT8 \\
Technology & TSMC 16nm & TSMC 16nm \\
Frequency (GHz) & 1.797 & 1.0 \\
Voltage (V) & 1.1 & 0.8 \\
Area (mm\textsuperscript{2}) & 3.1 & 1.8 \\
ResNet-50 Runtime (cycles) & 8.3M & 4.4M \\
ResNet-50 Runtime (ms) & 4.64 & 4.40 \\
MAC Utilization & 48\% & 95\% \\
\bottomrule
\end{tabular}
\caption{Comparison of Voyager with Simba.}
\label{tab:comparison_with_simba}
\end{table}

\begin{table}
\centering
\begin{tabular}{ccc}
\toprule
\textbf{Metric} & \textbf{4-bit VSQ~\cite{keller_transformer_jssc}} & \textbf{Voyager} \\
\midrule
\# MACs & 1024 & 1024 \\
Datatype & INT4 & INT4 \\
Technology & TSMC 5nm & TSMC 7nm \\
Frequency (MHz) & 909 & 909 \\
Voltage (V) & 0.67 & 0.75 \\
Area (mm\textsuperscript{2}) & 0.153 & 0.57 \\
BERT-Base Runtime (cycles) & 10.4M & 12.2M \\
BERT-Base Runtime (ms) & 11.4 & 13.4 \\
MAC Utilization & 98\% & 98\% \\
\bottomrule
\end{tabular}
\caption{Comparison of Voyager with 4-bit VSQ accelerator.}
\label{tab:comparison_with_vsq}
\end{table}
\end{comment}
\section{Conclusion}

Voyager is an end-to-end framework for generating DNN accelerators. Its HLS based accelerator generator provides extensive configurability in datatypes and quantization schemes. In addition, its PyTorch-based compiler maps models efficiently to generated hardware, supporting quantization, operation fusion, and tiling. 
Together, these components enable rapid design-space exploration, including the ability to evaluate quantized models on full datasets for precise accuracy estimation. 
Voyager-generated designs achieve consistently high utilization and performance across diverse models and at varying scales, significantly outperforming prior work.
Compared to prior generators, Voyager-generated designs achieve up to 56\% lower area and 61\% lower runtime at iso-hardware configurations. Against hand-optimized accelerators, Voyager achieves up to 48\% lower runtime, while requiring significantly less manual effort.  
We show that high performance DNN accelerators can be achieved without the extensive manual design and mapping  effort that has been traditionally required.

\section{Acknowledgments}
This work was supported by funding from DARPA Young Faculty Award, COCOSYS SRC JUMP 2.0 Center, Amazon, Apple, Siemens and PORTAL Affiliates Program.

\bibliographystyle{plain}
\bibliography{references}

\appendix
\section{Complete Runtime Results}

\autoref{tab:resnet18_complete}, \autoref{tab:resnet50_complete}, \autoref{tab:vit_complete}, \autoref{tab:bert_complete}, and \autoref{tab:mobilebert_complete} show detailed runtime results per layer for each reported model for 16$\times$16 and 32$\times$32 designs, to allow easy comparisons with related work. The total MAC utilization is calculated by summing ideal cycles for the matrix operations and summing achieved cycles for the matrix operations, and dividing ideal by achieved.

\onecolumn
\begin{longtable}{|l|c|c|r|r|r|r|r|r|}

\caption{ResNet-18 performance and MAC utilization per layer.} \label{tab:resnet18_complete} \\

\hline
Layer & Count & Unit &
\multicolumn{3}{c|}{16$\times$16} &
\multicolumn{3}{c|}{32$\times$32} \\
\cline{4-9}
& & & Ideal & Achieved & MAC & Ideal & Achieved & MAC \\
& & & Cycles & Cycles & Utilization & Cycles & Cycles & Utilization\\
\hline
\endhead

\hline
\multicolumn{9}{|r|}{Continued on next page...} \\
\hline
\endfoot

\hline
\endlastfoot

\hline
conv1\_fused & \multirow{2}{*}{1} & \multirow{2}{*}{matrix} & \multirow{2}{*}{460992} & \multirow{2}{*}{956615} & \multirow{2}{*}{48.19\%} & \multirow{2}{*}{115248} & \multirow{2}{*}{246291} & \multirow{2}{*}{46.79\%} \\
224$\times$224$\times$3 * 7$\times$7$\times$3$\times$64 & & & & & & & & \\
\hline
max\_pool2d\_default & \multirow{2}{*}{1} & \multirow{2}{*}{vector} & \multirow{2}{*}{--} & \multirow{2}{*}{223641} & \multirow{2}{*}{--} & \multirow{2}{*}{--} & \multirow{2}{*}{111859} & \multirow{2}{*}{--} \\
112$\times$112$\times$64 & & & & & & & & \\
\hline
quantize\_default\_1 & \multirow{2}{*}{1} & \multirow{2}{*}{vector} & \multirow{2}{*}{--} & \multirow{2}{*}{25163} & \multirow{2}{*}{--} & \multirow{2}{*}{--} & \multirow{2}{*}{12618} & \multirow{2}{*}{--} \\
56$\times$56$\times$64 & & & & & & & & \\
\hline
layer1\_0\_conv1\_fused & \multirow{2}{*}{2} & \multirow{2}{*}{matrix} & \multirow{2}{*}{451584} & \multirow{2}{*}{451850} & \multirow{2}{*}{99.94\%} & \multirow{2}{*}{112896} & \multirow{2}{*}{113366} & \multirow{2}{*}{99.59\%} \\
56$\times$56$\times$64 * 3$\times$3$\times$64$\times$64 & & & & & & & & \\
\hline
layer1\_0\_conv2\_fused & \multirow{2}{*}{2} & \multirow{2}{*}{matrix} & \multirow{2}{*}{451584} & \multirow{2}{*}{451850} & \multirow{2}{*}{99.94\%} & \multirow{2}{*}{112896} & \multirow{2}{*}{113366} & \multirow{2}{*}{99.59\%} \\
56$\times$56$\times$64 * 3$\times$3$\times$64$\times$64 & & & & & & & & \\
\hline
layer2\_0\_conv1\_fused & \multirow{2}{*}{1} & \multirow{2}{*}{matrix} & \multirow{2}{*}{225792} & \multirow{2}{*}{226378} & \multirow{2}{*}{99.74\%} & \multirow{2}{*}{56448} & \multirow{2}{*}{57518} & \multirow{2}{*}{98.14\%} \\
56$\times$56$\times$64 * 3$\times$3$\times$64$\times$128 & & & & & & & & \\
\hline
layer2\_0\_conv2\_fused & \multirow{2}{*}{1} & \multirow{2}{*}{matrix} & \multirow{2}{*}{451584} & \multirow{2}{*}{451870} & \multirow{2}{*}{99.94\%} & \multirow{2}{*}{112896} & \multirow{2}{*}{113355} & \multirow{2}{*}{99.59\%} \\
28$\times$28$\times$128 * 3$\times$3$\times$128$\times$128 & & & & & & & & \\
\hline
layer2\_0\_downsample\_0\_fused & \multirow{2}{*}{1} & \multirow{2}{*}{matrix} & \multirow{2}{*}{25088} & \multirow{2}{*}{25222} & \multirow{2}{*}{99.47\%} & \multirow{2}{*}{6272} & \multirow{2}{*}{6491} & \multirow{2}{*}{96.62\%} \\
56$\times$56$\times$64 * 1$\times$1$\times$64$\times$128 & & & & & & & & \\
\hline
layer2\_1\_conv1\_fused & \multirow{2}{*}{1} & \multirow{2}{*}{matrix} & \multirow{2}{*}{451584} & \multirow{2}{*}{451870} & \multirow{2}{*}{99.94\%} & \multirow{2}{*}{112896} & \multirow{2}{*}{113355} & \multirow{2}{*}{99.59\%} \\
28$\times$28$\times$128 * 3$\times$3$\times$128$\times$128 & & & & & & & & \\
\hline
layer2\_1\_conv2\_fused & \multirow{2}{*}{1} & \multirow{2}{*}{matrix} & \multirow{2}{*}{451584} & \multirow{2}{*}{451870} & \multirow{2}{*}{99.94\%} & \multirow{2}{*}{112896} & \multirow{2}{*}{113355} & \multirow{2}{*}{99.59\%} \\
28$\times$28$\times$128 * 3$\times$3$\times$128$\times$128 & & & & & & & & \\
\hline
layer3\_0\_conv1\_fused & \multirow{2}{*}{1} & \multirow{2}{*}{matrix} & \multirow{2}{*}{225792} & \multirow{2}{*}{226378} & \multirow{2}{*}{99.74\%} & \multirow{2}{*}{56448} & \multirow{2}{*}{57518} & \multirow{2}{*}{98.14\%} \\
28$\times$28$\times$128 * 3$\times$3$\times$128$\times$256 & & & & & & & & \\
\hline
layer3\_0\_conv2\_fused & \multirow{2}{*}{1} & \multirow{2}{*}{matrix} & \multirow{2}{*}{451584} & \multirow{2}{*}{451835} & \multirow{2}{*}{99.94\%} & \multirow{2}{*}{112896} & \multirow{2}{*}{113355} & \multirow{2}{*}{99.59\%} \\
14$\times$14$\times$256 * 3$\times$3$\times$256$\times$256 & & & & & & & & \\
\hline
layer3\_0\_downsample\_0\_fused & \multirow{2}{*}{1} & \multirow{2}{*}{matrix} & \multirow{2}{*}{25088} & \multirow{2}{*}{25222} & \multirow{2}{*}{99.47\%} & \multirow{2}{*}{6272} & \multirow{2}{*}{6491} & \multirow{2}{*}{96.62\%} \\
28$\times$28$\times$128 * 1$\times$1$\times$128$\times$256 & & & & & & & & \\
\hline
layer3\_1\_conv1\_fused & \multirow{2}{*}{1} & \multirow{2}{*}{matrix} & \multirow{2}{*}{451584} & \multirow{2}{*}{451835} & \multirow{2}{*}{99.94\%} & \multirow{2}{*}{112896} & \multirow{2}{*}{113355} & \multirow{2}{*}{99.59\%} \\
14$\times$14$\times$256 * 3$\times$3$\times$256$\times$256 & & & & & & & & \\
\hline
layer3\_1\_conv2\_fused & \multirow{2}{*}{1} & \multirow{2}{*}{matrix} & \multirow{2}{*}{451584} & \multirow{2}{*}{451835} & \multirow{2}{*}{99.94\%} & \multirow{2}{*}{112896} & \multirow{2}{*}{113355} & \multirow{2}{*}{99.59\%} \\
14$\times$14$\times$256 * 3$\times$3$\times$256$\times$256 & & & & & & & & \\
\hline
layer4\_0\_conv1\_fused & \multirow{2}{*}{1} & \multirow{2}{*}{matrix} & \multirow{2}{*}{225792} & \multirow{2}{*}{226154} & \multirow{2}{*}{99.84\%} & \multirow{2}{*}{56448} & \multirow{2}{*}{56907} & \multirow{2}{*}{99.19\%} \\
14$\times$14$\times$256 * 3$\times$3$\times$256$\times$512 & & & & & & & & \\
\hline
layer4\_0\_conv2\_fused & \multirow{2}{*}{1} & \multirow{2}{*}{matrix} & \multirow{2}{*}{451584} & \multirow{2}{*}{451835} & \multirow{2}{*}{99.94\%} & \multirow{2}{*}{112896} & \multirow{2}{*}{113355} & \multirow{2}{*}{99.59\%} \\
7$\times$7$\times$512 * 3$\times$3$\times$512$\times$512 & & & & & & & & \\
\hline
layer4\_0\_downsample\_0\_fused & \multirow{2}{*}{1} & \multirow{2}{*}{matrix} & \multirow{2}{*}{25088} & \multirow{2}{*}{25243} & \multirow{2}{*}{99.39\%} & \multirow{2}{*}{6272} & \multirow{2}{*}{6491} & \multirow{2}{*}{96.62\%} \\
14$\times$14$\times$256 * 1$\times$1$\times$256$\times$512 & & & & & & & & \\
\hline
layer4\_1\_conv1\_fused & \multirow{2}{*}{1} & \multirow{2}{*}{matrix} & \multirow{2}{*}{451584} & \multirow{2}{*}{451835} & \multirow{2}{*}{99.94\%} & \multirow{2}{*}{112896} & \multirow{2}{*}{113809} & \multirow{2}{*}{99.19\%} \\
7$\times$7$\times$512 * 3$\times$3$\times$512$\times$512 & & & & & & & & \\
\hline
layer4\_1\_conv2\_fused & \multirow{2}{*}{1} & \multirow{2}{*}{matrix} & \multirow{2}{*}{451584} & \multirow{2}{*}{452597} & \multirow{2}{*}{99.78\%} & \multirow{2}{*}{112896} & \multirow{2}{*}{113809} & \multirow{2}{*}{99.19\%} \\
7$\times$7$\times$512 * 3$\times$3$\times$512$\times$512 & & & & & & & & \\
\hline
adaptive\_avg\_pool2d\_default & \multirow{2}{*}{1} & \multirow{2}{*}{vector} & \multirow{2}{*}{--} & \multirow{2}{*}{3212} & \multirow{2}{*}{--} & \multirow{2}{*}{--} & \multirow{2}{*}{1643} & \multirow{2}{*}{--} \\
7$\times$7$\times$512 & & & & & & & & \\
\hline
fc & \multirow{2}{*}{1} & \multirow{2}{*}{vector} & \multirow{2}{*}{--} & \multirow{2}{*}{66038} & \multirow{2}{*}{--} & \multirow{2}{*}{--} & \multirow{2}{*}{33581} & \multirow{2}{*}{--} \\
1$\times$512 * 1000$\times$512 & & & & & & & & \\
\hline
\textbf{Total} &  &  & -- & 7904048 & 93.0\% & -- & 2071975 & 91.9\% \\

\end{longtable}

\begin{longtable}{|l|c|c|r|r|r|r|r|r|}
\caption{ResNet-50 performance and MAC utilization per layer.} \label{tab:resnet50_complete} \\
\hline
Layer & Count & Unit &
\multicolumn{3}{c|}{16$\times$16} &
\multicolumn{3}{c|}{32$\times$32} \\
\cline{4-9}
& & & Ideal & Achieved & MAC & Ideal & Achieved & MAC \\
& & & Cycles & Cycles & Utilization & Cycles & Cycles & Utilization\\
\hline
\endhead

\hline
\multicolumn{9}{|r|}{Continued on next page...} \\
\hline
\endfoot

\hline
\endlastfoot

\hline
quantize\_default\_1 & \multirow{2}{*}{1} & \multirow{2}{*}{vector} & \multirow{2}{*}{--} & \multirow{2}{*}{25163} & \multirow{2}{*}{--} & \multirow{2}{*}{--} & \multirow{2}{*}{12618} & \multirow{2}{*}{--} \\
56$\times$56$\times$64 & & & & & & & & \\
\hline
conv1\_fused & \multirow{2}{*}{1} & \multirow{2}{*}{matrix} & \multirow{2}{*}{460992} & \multirow{2}{*}{956615} & \multirow{2}{*}{48.19\%} & \multirow{2}{*}{115248} & \multirow{2}{*}{246291} & \multirow{2}{*}{46.79\%} \\
224$\times$224$\times$3 * 7$\times$7$\times$3$\times$64 & & & & & & & & \\
\hline
max\_pool2d\_default & \multirow{2}{*}{1} & \multirow{2}{*}{vector} & \multirow{2}{*}{--} & \multirow{2}{*}{223641} & \multirow{2}{*}{--} & \multirow{2}{*}{--} & \multirow{2}{*}{111859} & \multirow{2}{*}{--} \\
112$\times$112$\times$64 & & & & & & & & \\
\hline
layer1\_0\_conv1\_fused & \multirow{2}{*}{1} & \multirow{2}{*}{matrix} & \multirow{2}{*}{50176} & \multirow{2}{*}{50310} & \multirow{2}{*}{99.73\%} & \multirow{2}{*}{12544} & \multirow{2}{*}{12763} & \multirow{2}{*}{98.28\%} \\
56$\times$56$\times$64 * 1$\times$1$\times$64$\times$64 & & & & & & & & \\
\hline
layer1\_0\_conv2\_fused & \multirow{2}{*}{3} & \multirow{2}{*}{matrix} & \multirow{2}{*}{451584} & \multirow{2}{*}{451850} & \multirow{2}{*}{99.94\%} & \multirow{2}{*}{112896} & \multirow{2}{*}{113366} & \multirow{2}{*}{99.59\%} \\
56$\times$56$\times$64 * 3$\times$3$\times$64$\times$64 & & & & & & & & \\
\hline
layer1\_0\_conv3\_fused & \multirow{2}{*}{1} & \multirow{2}{*}{matrix} & \multirow{2}{*}{200704} & \multirow{2}{*}{200838} & \multirow{2}{*}{99.93\%} & \multirow{2}{*}{50176} & \multirow{2}{*}{50395} & \multirow{2}{*}{99.56\%} \\
56$\times$56$\times$64 * 1$\times$1$\times$64$\times$256 & & & & & & & & \\
\hline
layer1\_0\_downsample\_0\_fused & \multirow{2}{*}{1} & \multirow{2}{*}{matrix} & \multirow{2}{*}{200704} & \multirow{2}{*}{200838} & \multirow{2}{*}{99.93\%} & \multirow{2}{*}{50176} & \multirow{2}{*}{50395} & \multirow{2}{*}{99.56\%} \\
56$\times$56$\times$64 * 1$\times$1$\times$64$\times$256 & & & & & & & & \\
\hline
layer1\_1\_conv1\_fused & \multirow{2}{*}{2} & \multirow{2}{*}{matrix} & \multirow{2}{*}{200704} & \multirow{2}{*}{200838} & \multirow{2}{*}{99.93\%} & \multirow{2}{*}{50176} & \multirow{2}{*}{50395} & \multirow{2}{*}{99.56\%} \\
56$\times$56$\times$256 * 1$\times$1$\times$256$\times$64 & & & & & & & & \\
\hline
layer1\_1\_conv2\_fused & \multirow{2}{*}{2} & \multirow{2}{*}{matrix} & \multirow{2}{*}{451584} & \multirow{2}{*}{451870} & \multirow{2}{*}{99.94\%} & \multirow{2}{*}{112896} & \multirow{2}{*}{113355} & \multirow{2}{*}{99.59\%} \\
56$\times$56$\times$64 * 3$\times$3$\times$64$\times$64 & & & & & & & & \\
\hline
layer1\_1\_conv3\_fused & \multirow{2}{*}{3} & \multirow{2}{*}{matrix} & \multirow{2}{*}{200704} & \multirow{2}{*}{200838} & \multirow{2}{*}{99.93\%} & \multirow{2}{*}{50176} & \multirow{2}{*}{50395} & \multirow{2}{*}{99.56\%} \\
56$\times$56$\times$64 * 1$\times$1$\times$64$\times$256 & & & & & & & & \\
\hline
layer2\_0\_conv1\_fused & \multirow{2}{*}{1} & \multirow{2}{*}{matrix} & \multirow{2}{*}{401408} & \multirow{2}{*}{401542} & \multirow{2}{*}{99.97\%} & \multirow{2}{*}{100352} & \multirow{2}{*}{100571} & \multirow{2}{*}{99.78\%} \\
56$\times$56$\times$256 * 1$\times$1$\times$256$\times$128 & & & & & & & & \\
\hline
layer2\_0\_conv2\_fused & \multirow{2}{*}{1} & \multirow{2}{*}{matrix} & \multirow{2}{*}{451584} & \multirow{2}{*}{452170} & \multirow{2}{*}{99.87\%} & \multirow{2}{*}{112896} & \multirow{2}{*}{113966} & \multirow{2}{*}{99.06\%} \\
28$\times$28$\times$128 * 3$\times$3$\times$128$\times$128 & & & & & & & & \\
\hline
layer2\_0\_conv3\_fused & \multirow{2}{*}{1} & \multirow{2}{*}{matrix} & \multirow{2}{*}{200704} & \multirow{2}{*}{200838} & \multirow{2}{*}{99.93\%} & \multirow{2}{*}{50176} & \multirow{2}{*}{50395} & \multirow{2}{*}{99.56\%} \\
28$\times$28$\times$128 * 1$\times$1$\times$128$\times$512 & & & & & & & & \\
\hline
layer2\_0\_downsample\_0\_fused & \multirow{2}{*}{1} & \multirow{2}{*}{matrix} & \multirow{2}{*}{401408} & \multirow{2}{*}{401542} & \multirow{2}{*}{99.97\%} & \multirow{2}{*}{100352} & \multirow{2}{*}{100571} & \multirow{2}{*}{99.78\%} \\
56$\times$56$\times$256 * 1$\times$1$\times$256$\times$512 & & & & & & & & \\
\hline
layer2\_1\_conv1\_fused & \multirow{2}{*}{3} & \multirow{2}{*}{matrix} & \multirow{2}{*}{200704} & \multirow{2}{*}{200838} & \multirow{2}{*}{99.93\%} & \multirow{2}{*}{50176} & \multirow{2}{*}{50395} & \multirow{2}{*}{99.56\%} \\
28$\times$28$\times$512 * 1$\times$1$\times$512$\times$128 & & & & & & & & \\
\hline
layer2\_1\_conv2\_fused & \multirow{2}{*}{3} & \multirow{2}{*}{matrix} & \multirow{2}{*}{451584} & \multirow{2}{*}{451870} & \multirow{2}{*}{99.94\%} & \multirow{2}{*}{112896} & \multirow{2}{*}{113355} & \multirow{2}{*}{99.59\%} \\
28$\times$28$\times$128 * 3$\times$3$\times$128$\times$128 & & & & & & & & \\
\hline
layer2\_1\_conv3\_fused & \multirow{2}{*}{1} & \multirow{2}{*}{matrix} & \multirow{2}{*}{200704} & \multirow{2}{*}{200838} & \multirow{2}{*}{99.93\%} & \multirow{2}{*}{50176} & \multirow{2}{*}{50395} & \multirow{2}{*}{99.56\%} \\
28$\times$28$\times$128 * 1$\times$1$\times$128$\times$512 & & & & & & & & \\
\hline
layer3\_0\_conv1\_fused & \multirow{2}{*}{1} & \multirow{2}{*}{matrix} & \multirow{2}{*}{401408} & \multirow{2}{*}{401542} & \multirow{2}{*}{99.97\%} & \multirow{2}{*}{100352} & \multirow{2}{*}{100571} & \multirow{2}{*}{99.78\%} \\
28$\times$28$\times$512 * 1$\times$1$\times$512$\times$256 & & & & & & & & \\
\hline
layer3\_0\_conv2\_fused & \multirow{2}{*}{1} & \multirow{2}{*}{matrix} & \multirow{2}{*}{451584} & \multirow{2}{*}{452170} & \multirow{2}{*}{99.87\%} & \multirow{2}{*}{112896} & \multirow{2}{*}{113966} & \multirow{2}{*}{99.06\%} \\
14$\times$14$\times$256 * 3$\times$3$\times$256$\times$256 & & & & & & & & \\
\hline
layer3\_0\_conv3\_fused & \multirow{2}{*}{1} & \multirow{2}{*}{matrix} & \multirow{2}{*}{200704} & \multirow{2}{*}{200838} & \multirow{2}{*}{99.93\%} & \multirow{2}{*}{50176} & \multirow{2}{*}{50395} & \multirow{2}{*}{99.56\%} \\
14$\times$14$\times$256 * 1$\times$1$\times$256$\times$1024 & & & & & & & & \\
\hline
layer3\_0\_downsample\_0\_fused & \multirow{2}{*}{1} & \multirow{2}{*}{matrix} & \multirow{2}{*}{401408} & \multirow{2}{*}{401542} & \multirow{2}{*}{99.97\%} & \multirow{2}{*}{100352} & \multirow{2}{*}{100571} & \multirow{2}{*}{99.78\%} \\
28$\times$28$\times$512 * 1$\times$1$\times$512$\times$1024 & & & & & & & & \\
\hline
layer3\_1\_conv1\_fused & \multirow{2}{*}{5} & \multirow{2}{*}{matrix} & \multirow{2}{*}{200704} & \multirow{2}{*}{200838} & \multirow{2}{*}{99.93\%} & \multirow{2}{*}{50176} & \multirow{2}{*}{50395} & \multirow{2}{*}{99.56\%} \\
14$\times$14$\times$1024 * 1$\times$1$\times$1024$\times$256 & & & & & & & & \\
\hline
layer3\_1\_conv2\_fused & \multirow{2}{*}{5} & \multirow{2}{*}{matrix} & \multirow{2}{*}{451584} & \multirow{2}{*}{451835} & \multirow{2}{*}{99.94\%} & \multirow{2}{*}{112896} & \multirow{2}{*}{113355} & \multirow{2}{*}{99.59\%} \\
14$\times$14$\times$256 * 3$\times$3$\times$256$\times$256 & & & & & & & & \\
\hline
layer3\_1\_conv3\_fused & \multirow{2}{*}{1} & \multirow{2}{*}{matrix} & \multirow{2}{*}{200704} & \multirow{2}{*}{200838} & \multirow{2}{*}{99.93\%} & \multirow{2}{*}{50176} & \multirow{2}{*}{50395} & \multirow{2}{*}{99.56\%} \\
14$\times$14$\times$256 * 1$\times$1$\times$256$\times$1024 & & & & & & & & \\
\hline
layer4\_0\_conv1\_fused & \multirow{2}{*}{1} & \multirow{2}{*}{matrix} & \multirow{2}{*}{401408} & \multirow{2}{*}{401542} & \multirow{2}{*}{99.97\%} & \multirow{2}{*}{100352} & \multirow{2}{*}{100571} & \multirow{2}{*}{99.78\%} \\
14$\times$14$\times$1024 * 1$\times$1$\times$1024$\times$512 & & & & & & & & \\
\hline
layer4\_0\_conv2\_fused & \multirow{2}{*}{1} & \multirow{2}{*}{matrix} & \multirow{2}{*}{451584} & \multirow{2}{*}{451946} & \multirow{2}{*}{99.92\%} & \multirow{2}{*}{112896} & \multirow{2}{*}{113355} & \multirow{2}{*}{99.59\%} \\
7$\times$7$\times$512 * 3$\times$3$\times$512$\times$512 & & & & & & & & \\
\hline
layer4\_0\_conv3\_fused & \multirow{2}{*}{1} & \multirow{2}{*}{matrix} & \multirow{2}{*}{200704} & \multirow{2}{*}{200859} & \multirow{2}{*}{99.92\%} & \multirow{2}{*}{50176} & \multirow{2}{*}{50395} & \multirow{2}{*}{99.56\%} \\
7$\times$7$\times$512 * 1$\times$1$\times$512$\times$2048 & & & & & & & & \\
\hline
layer4\_0\_downsample\_0\_fused & \multirow{2}{*}{1} & \multirow{2}{*}{matrix} & \multirow{2}{*}{401408} & \multirow{2}{*}{401563} & \multirow{2}{*}{99.96\%} & \multirow{2}{*}{100352} & \multirow{2}{*}{100571} & \multirow{2}{*}{99.78\%} \\
14$\times$14$\times$1024 * 1$\times$1$\times$1024$\times$2048 & & & & & & & & \\
\hline
layer4\_1\_conv1\_fused & \multirow{2}{*}{2} & \multirow{2}{*}{matrix} & \multirow{2}{*}{200704} & \multirow{2}{*}{200859} & \multirow{2}{*}{99.92\%} & \multirow{2}{*}{50176} & \multirow{2}{*}{50395} & \multirow{2}{*}{99.56\%} \\
7$\times$7$\times$2048 * 1$\times$1$\times$2048$\times$512 & & & & & & & & \\
\hline
layer4\_1\_conv2\_fused & \multirow{2}{*}{2} & \multirow{2}{*}{matrix} & \multirow{2}{*}{451584} & \multirow{2}{*}{451835} & \multirow{2}{*}{99.94\%} & \multirow{2}{*}{112896} & \multirow{2}{*}{113355} & \multirow{2}{*}{99.59\%} \\
7$\times$7$\times$512 * 3$\times$3$\times$512$\times$512 & & & & & & & & \\
\hline
layer4\_1\_conv3\_fused & \multirow{2}{*}{1} & \multirow{2}{*}{matrix} & \multirow{2}{*}{200704} & \multirow{2}{*}{200859} & \multirow{2}{*}{99.92\%} & \multirow{2}{*}{50176} & \multirow{2}{*}{50395} & \multirow{2}{*}{99.56\%} \\
7$\times$7$\times$512 * 1$\times$1$\times$512$\times$2048 & & & & & & & & \\
\hline
layer4\_2\_conv3\_fused & \multirow{2}{*}{1} & \multirow{2}{*}{matrix} & \multirow{2}{*}{200704} & \multirow{2}{*}{205493} & \multirow{2}{*}{97.67\%} & \multirow{2}{*}{50176} & \multirow{2}{*}{52145} & \multirow{2}{*}{96.22\%} \\
7$\times$7$\times$512 * 1$\times$1$\times$512$\times$2048 & & & & & & & & \\
\hline
adaptive\_avg\_pool2d\_default & \multirow{2}{*}{1} & \multirow{2}{*}{vector} & \multirow{2}{*}{--} & \multirow{2}{*}{12620} & \multirow{2}{*}{--} & \multirow{2}{*}{--} & \multirow{2}{*}{6347} & \multirow{2}{*}{--} \\
7$\times$7$\times$2048 & & & & & & & & \\
\hline
fc & \multirow{2}{*}{1} & \multirow{2}{*}{vector} & \multirow{2}{*}{--} & \multirow{2}{*}{259574} & \multirow{2}{*}{--} & \multirow{2}{*}{--} & \multirow{2}{*}{131885} & \multirow{2}{*}{--} \\
1$\times$2048 * 1000$\times$2048 & & & & & & & & \\
\hline
\textbf{Total} &  &  & -- & 16896078 & 96.2\% & -- & 4378064 & 94.9\% \\

\end{longtable}

\vspace*{1\baselineskip}

\begin{longtable}{|l|c|c|r|r|r|r|r|r|}
\caption{ViT performance and MAC utilization per layer.} \label{tab:vit_complete} \\
\hline
Layer & Count & Unit &
\multicolumn{3}{c|}{16$\times$16} &
\multicolumn{3}{c|}{32$\times$32} \\
\cline{4-9}
& & & Ideal & Achieved & MAC & Ideal & Achieved & MAC \\
& & & Cycles & Cycles & Utilization & Cycles & Cycles & Utilization\\
\hline
\endhead

\hline
\multicolumn{9}{|r|}{Continued on next page...} \\
\hline
\endfoot

\hline
\endlastfoot

\hline
add & \multirow{2}{*}{1} & \multirow{2}{*}{vector} & \multirow{2}{*}{--} & \multirow{2}{*}{18989} & \multirow{2}{*}{--} & \multirow{2}{*}{--} & \multirow{2}{*}{9533} & \multirow{2}{*}{--} \\
197$\times$768 & & & & & & & & \\
\hline
add\_scalar & \multirow{2}{*}{1} & \multirow{2}{*}{vector} & \multirow{2}{*}{--} & \multirow{2}{*}{173} & \multirow{2}{*}{--} & \multirow{2}{*}{--} & \multirow{2}{*}{125} & \multirow{2}{*}{--} \\
1$\times$768 & & & & & & & & \\
\hline
classifier & \multirow{2}{*}{1} & \multirow{2}{*}{vector} & \multirow{2}{*}{--} & \multirow{2}{*}{98294} & \multirow{2}{*}{--} & \multirow{2}{*}{--} & \multirow{2}{*}{49965} & \multirow{2}{*}{--} \\
1$\times$768 * 768$\times$1000 & & & & & & & & \\
\hline
layer\_norm\_24 & \multirow{2}{*}{1} & \multirow{2}{*}{vector} & \multirow{2}{*}{--} & \multirow{2}{*}{86336} & \multirow{2}{*}{--} & \multirow{2}{*}{--} & \multirow{2}{*}{43326} & \multirow{2}{*}{--} \\
224$\times$768 & & & & & & & & \\
\hline
layer\_norm\_fused & \multirow{2}{*}{24} & \multirow{2}{*}{vector} & \multirow{2}{*}{--} & \multirow{2}{*}{86334} & \multirow{2}{*}{--} & \multirow{2}{*}{--} & \multirow{2}{*}{43323} & \multirow{2}{*}{--} \\
224$\times$768 & & & & & & & & \\
\hline
matmul\_24\_1\_fused & \multirow{2}{*}{144} & \multirow{2}{*}{matrix} & \multirow{2}{*}{12544} & \multirow{2}{*}{14500} & \multirow{2}{*}{86.51\%} & \multirow{2}{*}{3136} & \multirow{2}{*}{4439} & \multirow{2}{*}{70.65\%} \\
224$\times$64 * 64$\times$256 & & & & & & & & \\
\hline
matmul\_36\_1\_fused & \multirow{2}{*}{144} & \multirow{2}{*}{matrix} & \multirow{2}{*}{12544} & \multirow{2}{*}{12678} & \multirow{2}{*}{99.09\%} & \multirow{2}{*}{3136} & \multirow{2}{*}{3362} & \multirow{2}{*}{93.28\%} \\
224$\times$224 * 224$\times$64 & & & & & & & & \\
\hline
quantize\_default\_7 & \multirow{2}{*}{23} & \multirow{2}{*}{vector} & \multirow{2}{*}{--} & \multirow{2}{*}{21579} & \multirow{2}{*}{--} & \multirow{2}{*}{--} & \multirow{2}{*}{10826} & \multirow{2}{*}{--} \\
224$\times$768 & & & & & & & & \\
\hline
softmax\_12\_fused & \multirow{2}{*}{144} & \multirow{2}{*}{vector} & \multirow{2}{*}{--} & \multirow{2}{*}{19057} & \multirow{2}{*}{--} & \multirow{2}{*}{--} & \multirow{2}{*}{9646} & \multirow{2}{*}{--} \\
224$\times$256 & & & & & & & & \\
\hline
patch\_embeddings\_projection & \multirow{2}{*}{1} & \multirow{2}{*}{matrix} & \multirow{2}{*}{451584} & \multirow{2}{*}{603197} & \multirow{2}{*}{74.87\%} & \multirow{2}{*}{112896} & \multirow{2}{*}{172323} & \multirow{2}{*}{65.51\%} \\
224$\times$224$\times$3 * 16$\times$16$\times$3$\times$768 & & & & & & & & \\
\hline
attention\_key\_fused & \multirow{2}{*}{36} & \multirow{2}{*}{matrix} & \multirow{2}{*}{516096} & \multirow{2}{*}{516230} & \multirow{2}{*}{99.97\%} & \multirow{2}{*}{129024} & \multirow{2}{*}{129250} & \multirow{2}{*}{99.83\%} \\
224$\times$768 * 768$\times$768 & & & & & & & & \\
\hline
attention\_output\_dense\_fused\_0 & \multirow{2}{*}{1} & \multirow{2}{*}{matrix} & \multirow{2}{*}{516096} & \multirow{2}{*}{525447} & \multirow{2}{*}{98.22\%} & \multirow{2}{*}{129024} & \multirow{2}{*}{134528} & \multirow{2}{*}{95.91\%} \\
12$\times$224$\times$64 * 768$\times$768 & & & & & & & & \\
\hline
intermediate\_dense\_fused & \multirow{2}{*}{12} & \multirow{2}{*}{matrix} & \multirow{2}{*}{2064384} & \multirow{2}{*}{2064930} & \multirow{2}{*}{99.97\%} & \multirow{2}{*}{516096} & \multirow{2}{*}{519608} & \multirow{2}{*}{99.32\%} \\
224$\times$768 * 768$\times$3072 & & & & & & & & \\
\hline
output\_dense\_fused & \multirow{2}{*}{12} & \multirow{2}{*}{matrix} & \multirow{2}{*}{2064384} & \multirow{2}{*}{2064930} & \multirow{2}{*}{99.97\%} & \multirow{2}{*}{516096} & \multirow{2}{*}{519608} & \multirow{2}{*}{99.32\%} \\
224$\times$3072 * 3072$\times$768 & & & & & & & & \\
\hline
attention\_output\_dense\_fused\_1 & \multirow{2}{*}{11} & \multirow{2}{*}{matrix} & \multirow{2}{*}{516096} & \multirow{2}{*}{516641} & \multirow{2}{*}{99.90\%} & \multirow{2}{*}{129024} & \multirow{2}{*}{132536} & \multirow{2}{*}{97.35\%} \\
12$\times$224$\times$64 * 768$\times$768 & & & & & & & & \\
\hline
\textbf{Total} & & & -- & 84369325 & 99.3\% & -- & 22741874 & 98.0\% \\

\end{longtable}

\vspace*{1\baselineskip}

\begin{longtable}{|l|c|c|r|r|r|r|r|r|}
\caption{BERT-Base performance and MAC utilization per layer.} \label{tab:bert_complete} \\
\hline
Layer & Count & Unit &
\multicolumn{3}{c|}{16$\times$16} &
\multicolumn{3}{c|}{32$\times$32} \\
\cline{4-9}
& & & Ideal & Achieved & MAC & Ideal & Achieved & MAC \\
& & & Cycles & Cycles & Utilization & Cycles & Cycles & Utilization\\
\hline
\endhead

\hline
\multicolumn{9}{|r|}{Continued on next page...} \\
\hline
\endfoot

\hline
\endlastfoot

\hline
attention\_output\_dense\_fused & \multirow{2}{*}{12} & \multirow{2}{*}{matrix} & \multirow{2}{*}{294912} & \multirow{2}{*}{298537} & \multirow{2}{*}{98.79\%} & \multirow{2}{*}{73728} & \multirow{2}{*}{76485} & \multirow{2}{*}{96.40\%} \\
128$\times$768 * 768$\times$768 & & & & & & & & \\
\hline
attention\_self\_key\_fused & \multirow{2}{*}{36} & \multirow{2}{*}{matrix} & \multirow{2}{*}{294912} & \multirow{2}{*}{295053} & \multirow{2}{*}{99.95\%} & \multirow{2}{*}{73728} & \multirow{2}{*}{73965} & \multirow{2}{*}{99.68\%} \\
128$\times$768 * 768$\times$768 & & & & & & & & \\
\hline
intermediate\_dense\_fused & \multirow{2}{*}{12} & \multirow{2}{*}{matrix} & \multirow{2}{*}{1179648} & \multirow{2}{*}{1179790} & \multirow{2}{*}{99.99\%} & \multirow{2}{*}{294912} & \multirow{2}{*}{295149} & \multirow{2}{*}{99.92\%} \\
128$\times$768 * 768$\times$3072 & & & & & & & & \\
\hline
output\_dense\_fused & \multirow{2}{*}{12} & \multirow{2}{*}{matrix} & \multirow{2}{*}{1179648} & \multirow{2}{*}{1180393} & \multirow{2}{*}{99.94\%} & \multirow{2}{*}{294912} & \multirow{2}{*}{297092} & \multirow{2}{*}{99.26\%} \\
128$\times$3072 * 3072$\times$768 & & & & & & & & \\
\hline
classifier & \multirow{2}{*}{1} & \multirow{2}{*}{vector} & \multirow{2}{*}{--} & \multirow{2}{*}{1635} & \multirow{2}{*}{--} & \multirow{2}{*}{--} & \multirow{2}{*}{1634} & \multirow{2}{*}{--} \\
1$\times$768 * 768$\times$2 & & & & & & & & \\
\hline
layer\_norm\_23 & \multirow{2}{*}{1} & \multirow{2}{*}{vector} & \multirow{2}{*}{--} & \multirow{2}{*}{49464} & \multirow{2}{*}{--} & \multirow{2}{*}{3072} & \multirow{2}{*}{24886} & \multirow{2}{*}{--} \\
128$\times$768 & & & & & & & & \\
\hline
layer\_norm\_fused & \multirow{2}{*}{23} & \multirow{2}{*}{vector} & \multirow{2}{*}{--} & \multirow{2}{*}{49462} & \multirow{2}{*}{--} & \multirow{2}{*}{3072} & \multirow{2}{*}{24883} & \multirow{2}{*}{--} \\
128$\times$768 & & & & & & & & \\
\hline
matmul\_24\_fused & \multirow{2}{*}{144} & \multirow{2}{*}{matrix} & \multirow{2}{*}{4096} & \multirow{2}{*}{5135} & \multirow{2}{*}{79.77\%} & \multirow{2}{*}{1024} & \multirow{2}{*}{1778} & \multirow{2}{*}{57.57\%} \\
128$\times$64 * 64$\times$128 & & & & & & & & \\
\hline
matmul\_36\_fused & \multirow{2}{*}{144} & \multirow{2}{*}{matrix} & \multirow{2}{*}{4096} & \multirow{2}{*}{4237} & \multirow{2}{*}{96.67\%} & \multirow{2}{*}{1024} & \multirow{2}{*}{1261} & \multirow{2}{*}{81.18\%} \\
128$\times$128 * 128$\times$64 & & & & & & & & \\
\hline
pooler\_dense & \multirow{2}{*}{1} & \multirow{2}{*}{matrix} & \multirow{2}{*}{36864} & \multirow{2}{*}{74955} & \multirow{2}{*}{49.17\%} & \multirow{2}{*}{18432} & \multirow{2}{*}{37491} & \multirow{2}{*}{49.15\%} \\
1$\times$768 * 768$\times$768 & & & & & & & & \\
\hline
quantize\_default & \multirow{2}{*}{1} & \multirow{2}{*}{vector} & \multirow{2}{*}{--} & \multirow{2}{*}{12361} & \multirow{2}{*}{--} & \multirow{2}{*}{--} & \multirow{2}{*}{6216} & \multirow{2}{*}{--} \\
128$\times$768 & & & & & & & & \\
\hline
softmax\_12\_fused & \multirow{2}{*}{144} & \multirow{2}{*}{vector} & \multirow{2}{*}{--} & \multirow{2}{*}{6379} & \multirow{2}{*}{--} & \multirow{2}{*}{--} & \multirow{2}{*}{3304} & \multirow{2}{*}{--} \\
128$\times$128 & & & & & & & & \\
\hline
tanh & \multirow{2}{*}{1} & \multirow{2}{*}{vector} & \multirow{2}{*}{--} & \multirow{2}{*}{171} & \multirow{2}{*}{--} & \multirow{2}{*}{--} & \multirow{2}{*}{123} & \multirow{2}{*}{--} \\
1$\times$768 & & & & & & & & \\
\hline
\hline
\textbf{Total} & & & -- & 46070904 & 99.4\% & -- & 12243503 & 97.9\% \\

\end{longtable}

\begin{longtable}{|l|c|c|r|r|r|r|r|r|}
\caption{MobileBERT-TINY performance and MAC utilization per layer.} \label{tab:mobilebert_complete} \\
\hline
Layer & Count & Unit &
\multicolumn{3}{c|}{16$\times$16} &
\multicolumn{3}{c|}{32$\times$32} \\
\cline{4-9}
& & & Ideal & Achieved & MAC & Ideal & Achieved & MAC \\
& & & Cycles & Cycles & Utilization & Cycles & Cycles & Utilization\\\hline
\endhead

\hline
\multicolumn{9}{|r|}{Continued on next page...} \\
\hline
\endfoot

\hline
\endlastfoot

\hline
add\_10 & \multirow{2}{*}{21} & \multirow{2}{*}{vector} & \multirow{2}{*}{--} & \multirow{2}{*}{8267} & \multirow{2}{*}{--} & \multirow{2}{*}{--} & \multirow{2}{*}{4171} & \multirow{2}{*}{--} \\
128$\times$512 & & & & & & & & \\
\hline
add\_4\_fused & \multirow{2}{*}{63} & \multirow{2}{*}{vector} & \multirow{2}{*}{--} & \multirow{2}{*}{2121} & \multirow{2}{*}{--} & \multirow{2}{*}{--} & \multirow{2}{*}{1096} & \multirow{2}{*}{--} \\
128$\times$128 & & & & & & & & \\
\hline
classifier & \multirow{2}{*}{21} & \multirow{2}{*}{vector} & \multirow{2}{*}{--} & \multirow{2}{*}{1123} & \multirow{2}{*}{--} & \multirow{2}{*}{--} & \multirow{2}{*}{1122} & \multirow{2}{*}{--} \\
1$\times$512 * 512$\times$2 & & & & & & & & \\
\hline
matmul\_2\_fused & \multirow{2}{*}{84} & \multirow{2}{*}{matrix} & \multirow{2}{*}{2048} & \multirow{2}{*}{3094} & \multirow{2}{*}{66.18\%} & \multirow{2}{*}{512} & \multirow{2}{*}{1262} & \multirow{2}{*}{40.57\%} \\
128$\times$32 * 32$\times$128 & & & & & & & & \\
\hline
matmul\_6\_fused & \multirow{2}{*}{84} & \multirow{2}{*}{matrix} & \multirow{2}{*}{2048} & \multirow{2}{*}{2189} & \multirow{2}{*}{93.56\%} & \multirow{2}{*}{512} & \multirow{2}{*}{748} & \multirow{2}{*}{68.45\%} \\
128$\times$128 * 128$\times$32 & & & & & & & & \\
\hline
attention\_output\_dense\_fused & \multirow{2}{*}{21} & \multirow{2}{*}{matrix} & \multirow{2}{*}{8192} & \multirow{2}{*}{9294} & \multirow{2}{*}{88.12\%} & \multirow{2}{*}{2048} & \multirow{2}{*}{2791} & \multirow{2}{*}{73.38\%} \\
4$\times$128$\times$32 * 128$\times$128 & & & & & & & & \\
\hline
attention\_self\_query\_fused & \multirow{2}{*}{42} & \multirow{2}{*}{matrix} & \multirow{2}{*}{8192} & \multirow{2}{*}{8333} & \multirow{2}{*}{98.31\%} & \multirow{2}{*}{2048} & \multirow{2}{*}{2285} & \multirow{2}{*}{89.60\%} \\
128$\times$128 * 128$\times$128 & & & & & & & & \\
\hline
attention\_self\_value\_fused & \multirow{2}{*}{21} & \multirow{2}{*}{matrix} & \multirow{2}{*}{32768} & \multirow{2}{*}{32908} & \multirow{2}{*}{99.57\%} & \multirow{2}{*}{8192} & \multirow{2}{*}{8428} & \multirow{2}{*}{97.19\%} \\
128$\times$512 * 512$\times$128 & & & & & & & & \\
\hline
bottleneck\_input\_dense\_fused & \multirow{2}{*}{42} & \multirow{2}{*}{matrix} & \multirow{2}{*}{32768} & \multirow{2}{*}{33802} & \multirow{2}{*}{96.94\%} & \multirow{2}{*}{8192} & \multirow{2}{*}{8926} & \multirow{2}{*}{91.78\%} \\
128$\times$512 * 512$\times$128 & & & & & & & & \\
\hline
ffn\_intermediate\_dense\_fused & \multirow{2}{*}{42} & \multirow{2}{*}{matrix} & \multirow{2}{*}{32768} & \multirow{2}{*}{32909} & \multirow{2}{*}{99.57\%} & \multirow{2}{*}{8192} & \multirow{2}{*}{8429} & \multirow{2}{*}{97.18\%} \\
128$\times$128 * 128$\times$512 & & & & & & & & \\
\hline
ffn\_output\_dense\_fused & \multirow{2}{*}{42} & \multirow{2}{*}{matrix} & \multirow{2}{*}{32768} & \multirow{2}{*}{33805} & \multirow{2}{*}{96.93\%} & \multirow{2}{*}{8192} & \multirow{2}{*}{8930} & \multirow{2}{*}{91.77\%} \\
128$\times$512 * 512$\times$128 & & & & & & & & \\
\hline
output\_bottleneck\_dense\_fused & \multirow{2}{*}{21} & \multirow{2}{*}{matrix} & \multirow{2}{*}{32768} & \multirow{2}{*}{36750} & \multirow{2}{*}{89.16\%} & \multirow{2}{*}{8192} & \multirow{2}{*}{10459} & \multirow{2}{*}{78.32\%} \\
128$\times$128 * 128$\times$512 & & & & & & & & \\
\hline
quantize\_default & \multirow{2}{*}{21} & \multirow{2}{*}{vector} & \multirow{2}{*}{--} & \multirow{2}{*}{8265} & \multirow{2}{*}{--} & \multirow{2}{*}{--} & \multirow{2}{*}{4168} & \multirow{2}{*}{--} \\
128$\times$512 & & & & & & & & \\
\hline
softmax\_1\_fused & \multirow{2}{*}{84} & \multirow{2}{*}{vector} & \multirow{2}{*}{--} & \multirow{2}{*}{6379} & \multirow{2}{*}{--} & \multirow{2}{*}{--} & \multirow{2}{*}{3304} & \multirow{2}{*}{--} \\
128$\times$128 & & & & & & & & \\
\hline
\textbf{Total} & & & -- & 7713636 & 95.1\% & -- & 2369283 & 86.2\% \\

\end{longtable}

\begin{longtable}{|l|c|c|r|r|r|}
\caption{LLaMA performance and MAC utilization per layer.} \label{tab:llama_complete} \\
\hline
Layer & Count & Unit &
\multicolumn{3}{c|}{32$\times$32} \\
\cline{4-6}
& & & Ideal & Achieved & MAC \\
& & & Cycles & Cycles & Utilization \\\hline
\endhead

\hline
\multicolumn{6}{|r|}{Continued on next page...} \\
\hline
\endfoot

\hline
\endlastfoot

\hline
add\_39\_fused & \multirow{2}{*}{128} & \multirow{2}{*}{vector} & \multirow{2}{*}{--} & \multirow{2}{*}{16,456} & \multirow{2}{*}{--} \\
1$\times$8$\times$512$\times$64 & & & & & \\
\hline
layernorm\_default\_fused & \multirow{2}{*}{32} & \multirow{2}{*}{vector} & \multirow{2}{*}{--} & \multirow{2}{*}{262,451} & \multirow{2}{*}{--} \\
1$\times$32$\times$64$\times$512 & & & & & \\

\hline
matmul\_2\_fused & \multirow{2}{*}{512} & \multirow{2}{*}{matrix} & \multirow{2}{*}{16384} & \multirow{2}{*}{16754} & \multirow{2}{*}{97.79\%} \\
512$\times$512 * 512$\times$64 & & & & & \\
\hline
matmul\_34\_fused & \multirow{2}{*}{512} & \multirow{2}{*}{matrix} & \multirow{2}{*}{16384} & \multirow{2}{*}{16620} & \multirow{2}{*}{98.58\%} \\
512$\times$64 * 64$\times$512 & & & & & \\

\hline
model\_layers\_0\_mlp\_down\_proj\_fused & \multirow{2}{*}{16} & \multirow{2}{*}{matrix} & \multirow{2}{*}{8,388,608} & \multirow{2}{*}{8,390,496} & \multirow{2}{*}{99.98\%} \\
512$\times$8192 * 8192$\times$2048 & & & & & \\
\hline
model\_layers\_0\_mlp\_gate\_proj\_fused & \multirow{2}{*}{16} & \multirow{2}{*}{matrix} & \multirow{2}{*}{16,777,216} & \multirow{2}{*}{16,780,160} & \multirow{2}{*}{99.98\%} \\
512$\times$2048 * 2048$\times$8192 & & & & & \\
\hline
model\_layers\_0\_mlp\_up\_proj\_fused & \multirow{2}{*}{16} & \multirow{2}{*}{matrix} & \multirow{2}{*}{8,388,608} & \multirow{2}{*}{8,391,520} & \multirow{2}{*}{99.79\%} \\
512$\times$2048 * 2048$\times$8192 & & & & & \\

\hline
model\_layers\_0\_self\_attn\_k\_proj\_fused & \multirow{2}{*}{32} & \multirow{2}{*}{matrix} & \multirow{2}{*}{524,288} & \multirow{2}{*}{524,656} & \multirow{2}{*}{99.93\%} \\
512$\times$2048 * 2048$\times$512 & & & & & \\
\hline
model\_layers\_0\_self\_attn\_o\_proj\_fused & \multirow{2}{*}{16} & \multirow{2}{*}{matrix} & \multirow{2}{*}{2,097,152} & \multirow{2}{*}{2,097,520} & \multirow{2}{*}{99.98\%} \\
512$\times$2048 * 2048$\times$2048 & & & & & \\
\hline
model\_layers\_0\_self\_attn\_q\_proj\_fused & \multirow{2}{*}{16} & \multirow{2}{*}{matrix} & \multirow{2}{*}{2,097,152} & \multirow{2}{*}{2,097,520} & \multirow{2}{*}{99.98\%} \\
512$\times$2048 * 2048$\times$2048 & & & & & \\

\hline
mul\_2\_fused & \multirow{2}{*}{16} & \multirow{2}{*}{vector} & \multirow{2}{*}{--} & \multirow{2}{*}{65608} & \multirow{2}{*}{--} \\
512$\times$2048 * 2048$\times$2048 & & & & & \\
\hline
mul\_4\_fused & \multirow{2}{*}{16} & \multirow{2}{*}{vector} & \multirow{2}{*}{--} & \multirow{2}{*}{16456} & \multirow{2}{*}{--} \\
512$\times$2048 * 2048$\times$2048 & & & & & \\

\hline
permute\_default\_1\_fused & \multirow{2}{*}{16} & \multirow{2}{*}{vector} & \multirow{2}{*}{--} & \multirow{2}{*}{16459} & \multirow{2}{*}{--} \\
4$\times$1$\times$8$\times$512$\times$64 & & & & & \\
\hline
permute\_default\_fused & \multirow{2}{*}{16} & \multirow{2}{*}{vector} & \multirow{2}{*}{--} & \multirow{2}{*}{65608} & \multirow{2}{*}{--} \\
512$\times$2048 * 2048$\times$2048 & & & & & \\
\hline
quantize\_default & \multirow{2}{*}{32} & \multirow{2}{*}{vector} & \multirow{2}{*}{--} & \multirow{2}{*}{65608} & \multirow{2}{*}{--} \\
32$\times$64$\times$512 & & & & & \\

\hline
slice\_1 & \multirow{2}{*}{16} & \multirow{2}{*}{vector} & \multirow{2}{*}{--} & \multirow{2}{*}{32843} & \multirow{2}{*}{--} \\
32$\times$512$\times$64 -> 32$\times$64$\times$512 & & & & & \\
\hline
slice\_2\_fused & \multirow{2}{*}{16} & \multirow{2}{*}{vector} & \multirow{2}{*}{--} & \multirow{2}{*}{32843} & \multirow{2}{*}{--} \\
32$\times$512$\times$64 -> 512$\times$32$\times$64 & & & & & \\
\hline
slice\_3 & \multirow{2}{*}{16} & \multirow{2}{*}{vector} & \multirow{2}{*}{--} & \multirow{2}{*}{8267} & \multirow{2}{*}{--} \\
32$\times$512$\times$64 -> 32$\times$64$\times$512 & & & & & \\
\hline
slice\_4\_fused & \multirow{2}{*}{16} & \multirow{2}{*}{vector} & \multirow{2}{*}{--} & \multirow{2}{*}{8267} & \multirow{2}{*}{--} \\
32$\times$512$\times$64 -> 512$\times$32$\times$64 & & & & & \\
\hline
softmax\_1\_fused & \multirow{2}{*}{512} & \multirow{2}{*}{vector} & \multirow{2}{*}{--} & \multirow{2}{*}{49384} & \multirow{2}{*}{--} \\
512$\times$512 & & & & & \\

\hline
lm\_head & \multirow{2}{*}{1} & \multirow{2}{*}{matrix} & \multirow{2}{*}{8,126,464} & \multirow{2}{*}{8,127,936} & \multirow{2}{*}{99.98\%} \\
512$\times$2048 * 2048$\times$126976 & & & & & \\

\hline
\textbf{Total} & & & -- & 687,950,352 & 99.8\% \\

\end{longtable}

\begin{longtable}{|l|c|c|r|r|r|r|r|r|}
\caption{MobileNetV2 performance and MAC utilization per layer.} \label{tab:mobilenetv2_complete} \\
\hline
Layer & Count & Unit &
\multicolumn{3}{c|}{16$\times$16} &
\multicolumn{3}{c|}{32$\times$32} \\
\cline{4-9}
& & & Ideal & Achieved & MAC & Ideal & Achieved & MAC \\
& & & Cycles & Cycles & Utilization & Cycles & Cycles & Utilization\\\hline
\endhead

\hline
\multicolumn{9}{|r|}{Continued on next page...} \\
\hline
\endfoot

\hline
\endlastfoot

\hline
features\_0\_0\_fused & 1 & matrix & 42336 & 466179 & 9.08\% & 10584 & 117505 & 9.00\% \\
224$\times$224$\times$3 * 3$\times$3$\times$3$\times$32 & & & & & & & & \\
\hline
features\_1\_conv\_0\_0\_fused & 1 & dwc & 25088 & 26794 & 93.63\% & 12544 & 13460 & 93.19\% \\
112$\times$112$\times$32 * 3$\times$3$\times$1$\times$32 & & & & & & & & \\
\hline
features\_1\_conv\_1\_fused & 1 & matrix & 25088 & 25224 & 99.46\% & 6272& 12765 & 49.13\% \\
112$\times$112$\times$16 * 1$\times$1$\times$32$\times$16 & & & & & & & & \\
\hline
features\_2\_conv\_0\_0\_fused & 1 & matrix & 75264 & 75400 & 99.82\% & 18816 & 37853 & 49.71\% \\
112$\times$112$\times$16 * 1$\times$1$\times$16$\times$96 & & & & & & & & \\
\hline
features\_2\_conv\_1\_0\_fused & 1 & dwc & 18816 & 80155 & 23.47\% & 9408 & 40153 & 23.43\% \\
112$\times$112$\times$96 * 3$\times$3$\times$1$\times$96 & & & & & & & & \\
\hline
features\_2\_conv\_2\_fused & 1 & matrix & 28224 & 37768 & 74.73\% & 7056 & 9629 & 73.28\% \\
56$\times$56$\times$96 * 1$\times$1$\times$96$\times$24 & & & & & & & & \\
\hline
features\_3\_conv\_0\_0\_fused & 2 & matrix & 42336 & 56584 & 74.82\% & 10584 & 15901 & 66.56\% \\
56$\times$56$\times$24 * 1$\times$1$\times$24$\times$144 & & & & & & & & \\
\hline
features\_3\_conv\_1\_0\_fused & 1 & dwc & 28224 & 31018 & 90.99\% & 14112 & 17354 & 81.32\% \\
56$\times$56$\times$144 * 3$\times$3$\times$1$\times$144 & & & & & & & & \\
\hline
features\_3\_conv\_2\_fused & 1 & matrix & 42336 & 56584 & 74.82\% & 10584 & 15901 & 66.56\% \\
56$\times$56$\times$144 * 1$\times$1$\times$144$\times$24 & & & & & & & & \\
\hline
features\_4\_conv\_1\_0\_fused & 1 & dwc & 7056 & 30997 & 22.76\% & 3528 & 17333 & 20.35\% \\
56$\times$56$\times$144 * 3$\times$3$\times$1$\times$144 & & & & & & & & \\
\hline
features\_4\_conv\_2\_fused & 1 & matrix & 14112 & 14248 & 99.05\% & 3528 & 4141 & 85.20\% \\
28$\times$28$\times$144 * 1$\times$1$\times$144$\times$32 & & & & & & & & \\
\hline
features\_5\_conv\_0\_0\_fused & 3 & matrix & 18816 & 18952 & 99.28\% & 4704 & 4925 & 95.51\% \\
28$\times$28$\times$32 * 1$\times$1$\times$32$\times$192 & & & & & & & & \\
\hline
features\_5\_conv\_1\_0\_fused & 2 & dwc & 9408 & 10726 & 87.71\% & 4704 & 5506 & 85.43\% \\
28$\times$28$\times$192 * 3$\times$3$\times$1$\times$192 & & & & & & & & \\
\hline
features\_5\_conv\_2\_fused & 2 & matrix & 18816 & 18952 & 99.28\% & 4704 & 4925 & 95.51\% \\
28$\times$28$\times$192 * 1$\times$1$\times$192$\times$32 & & & & & & & & \\
\hline
features\_7\_conv\_1\_0\_fused & 1 & dwc & 2352 & 10695 & 21.99\% & 1176 & 5475 & 21.48\% \\
28$\times$28$\times$192 * 3$\times$3$\times$1$\times$192 & & & & & & & & \\
\hline
features\_7\_conv\_2\_fused & 1 & matrix & 9408 & 9544 & 98.58\% & 2352 & 2573 & 91.41\% \\
14$\times$14$\times$192 * 1$\times$1$\times$192$\times$64 & & & & & & & & \\
\hline
features\_8\_conv\_0\_0\_fused & 4 & matrix & 18816 & 18952 & 99.28\% & 4704 & 4925 & 95.51\% \\
14$\times$14$\times$64 * 1$\times$1$\times$64$\times$384 & & & & & & & & \\
\hline
features\_8\_conv\_1\_0\_fused & 4 & dwc & 4704 & 6238 & 75.41\% & 2352 & 3358 & 70.04\% \\
14$\times$14$\times$384 * 3$\times$3$\times$1$\times$384 & & & & & & & & \\
\hline
features\_8\_conv\_2\_fused & 3 & matrix & 18816 & 18952 & 99.28\% & 4704 & 4925 & 95.51\% \\
14$\times$14$\times$384 * 1$\times$1$\times$384$\times$64 & & & & & & & & \\
\hline
features\_11\_conv\_2\_fused & 1 & matrix & 28224 & 28360 & 99.52\% & 7056 & 7277 & 96.96\% \\
14$\times$14$\times$384 * 1$\times$1$\times$384$\times$96 & & & & & & & & \\
\hline
features\_12\_conv\_0\_0\_fused & 3 & matrix & 42336 & 42472 & 99.68\% & 10584 & 10805 & 97.95\% \\
14$\times$14$\times$96 * 1$\times$1$\times$96$\times$576 & & & & & & & & \\
\hline
features\_12\_conv\_1\_0\_fused & 2 & dwc & 7056 & 9310 & 75.79\% & 3528 & 4990 & 70.70\% \\
14$\times$14$\times$576 * 3$\times$3$\times$1$\times$576 & & & & & & & & \\
\hline
features\_12\_conv\_2\_fused & 2 & matrix & 42336 & 42472 & 99.68\% & 10584 & 10805 & 97.95\% \\
14$\times$14$\times$576 * 1$\times$1$\times$576$\times$96 & & & & & & & & \\
\hline
features\_14\_conv\_1\_0\_fused & 1 & dwc & 1764 & 9293 & 18.98\% & 882 & 4973 & 17.73\% \\
14$\times$14$\times$576 * 3$\times$3$\times$1$\times$576 & & & & & & & & \\
\hline
features\_14\_conv\_2\_fused & 1 & matrix & 17640 & 17797 & 99.12\% & 4410 & 4631 & 95.23\% \\
7$\times$7$\times$576 * 1$\times$1$\times$576$\times$160 & & & & & & & & \\
\hline
features\_15\_conv\_0\_0\_fused & 3 & matrix & 29400 & 29557 & 99.47\% & 7350 & 7571 & 97.08\% \\
7$\times$7$\times$160 * 1$\times$1$\times$160$\times$960 & & & & & & & & \\
\hline
features\_15\_conv\_1\_0\_fused & 3 & dwc & 2940 & 5374 & 54.71\% & 1470 & 3214 & 45.74\% \\
7$\times$7$\times$960 * 3$\times$3$\times$1$\times$960 & & & & & & & & \\
\hline
features\_15\_conv\_2\_fused & 2 & matrix & 29400 & 29557 & 99.47\% & 7350 & 7571 & 97.08\% \\
7$\times$7$\times$960 * 1$\times$1$\times$960$\times$160 & & & & & & & & \\
\hline
features\_17\_conv\_2\_fused & 1 & matrix & 58800 & 58957 & 99.73\% & 14700 & 14921 & 98.52\% \\
7$\times$7$\times$960 * 1$\times$1$\times$960$\times$320 & & & & & & & & \\
\hline
features\_18\_0\_fused & 1 & matrix & 78400 & 78659 & 99.67\% & 19600 & 19923 & 98.38\% \\
7$\times$7$\times$320 * 1$\times$1$\times$320$\times$1280 & & & & & & & & \\
\hline
adaptive\_avg\_pool2d\_default & 1 & vector & -- & 7914 & -- & -- & 3993 & -- \\
7$\times$7$\times$1280 & & & & & & & & \\
\hline
classifier\_1 & 1 & vector & -- & 162867 & -- & -- & 82731 & -- \\
1280 * 1280$\times$1000 & & & & & & & & \\
\hline

\textbf{Total} & & & -- & 2010336 & 83.7\% & -- & 659439 & 78.0\% \\

\end{longtable}

\twocolumn

% \section{Additional Comparisons}

% Table \ref{tab:comparison_with_snap} compares Voyager with SNAP~\cite{snap}, a sparse accelerator for CNNs. Sparse accelerators require complex interconnects for dynamic dataflows, leading to large area and lower frequencies, with SNAP consuming over 2$\times$ the area of the Voyager-generated design.

% \begin{table}
% \centering
% \begin{tabular}{ccc}
% \toprule
% \textbf{Metric} & \textbf{SNAP~\cite{snap}} & \textbf{Voyager} \\
% \midrule
% \# MACs & 252 & 256 \\
% Datatype & INT16 & INT16 \\
% Technology & 16nm & 16nm \\
% Frequency (MHz) & 480 & 480 \\
% Voltage (V) & 0.8 & 0.8 \\
% Area (mm\textsuperscript{2}) & 2.4 & 1.1 \\
% ResNet-50 cycles & 5.28M & 17.0M \\
% ResNet-50 ms & 10.99 & 35.56 \\
% MAC Utilization & -- & 96.2\% \\
% \bottomrule
% \end{tabular}
% \caption{Comparison of Voyager with a sparse accelerator.}
% \label{tab:comparison_with_snap}
% \end{table}

\end{document}